\documentclass[12pt]{article}
\usepackage{graphicx}
\usepackage{amsfonts}
\usepackage{amssymb}
\usepackage{mathrsfs}
\usepackage{amsmath}
\usepackage[unicode]{hyperref}
\usepackage{booktabs}
\begin{document}
\renewcommand{\thefootnote}{\fnsymbol{footnote}}
\begin{titlepage}

\vspace{10mm}
\begin{center}
{\Large\bf Three-generation neutrino oscillations in curved spacetime}
\vspace{8mm}

{\large Yu-Hao Zhang${}^{1,}$\footnote{E-mail address: yhzhang1994@gmail.com} and Xue-Qian Li${}^{1,}$\footnote{E-mail address: lixq@nankai.edu.cn}

\vspace{6mm}
${}^{1}${\normalsize \em School of Physics, Nankai University, Tianjin 300071, China}

}

\end{center}

\vspace{4mm}
\centerline{{\bf{Abstract}}}
\vspace{4mm}
Three-generation MSW effect in curved spacetime is studied and a brief discussion on the gravitational correction to the neutrino self-energy is given. The modified mixing parameters and corresponding conversion probabilities of neutrinos after traveling through celestial objects of constant densities are obtained. The method to distinguish between the normal hierarchy and inverted hierarchy is discussed in this framework. Due to the gravitational redshift of energy, in some extreme situations, the resonance energy of neutrinos might be shifted noticeably and the gravitational effect on the self-energy of neutrino becomes significant at the vicinities of spacetime singularities.

%

\end{titlepage}
\newpage
\renewcommand{\thefootnote}{\arabic{footnote}}
\setcounter{footnote}{0}
\setcounter{page}{2}
\section{Introduction}
The theory about neutrino has so far been well-established and oscillations among three generations are experimentally observed as the mixing angles are measured with rather high
accuracy based on the Earth experiments with long and short baselines, even though the CP violating phase is not well determined yet. In terms of the measured values the solar and atmospheric neutrino phenomena are perfectly explained. Now people turn their attention to the cosmic neutrino which, as is well known, serves as a unique messenger to provide us with valuable information of properties of super-large celestial objects, explosions of supernovae and even the structure of the earlier universe which was not optically transparent. This topic attracts much attention from theorists and experimentalists of particle physics. On the long way from the production source to the earth detectors, the neutrino beam would pass many  star clusters and galaxies (including the source of production) whose large mass would curve the spacetime of their vicinities. As it traverses the massive region, it undergoes the Mikheyev-Smirnov-Wolfenstein (MSW) effect\cite{wolfenstein78} which determines its evolution. It is well understood how the electron neutrino produced in the Sun converses to other flavors due to the MSW effect. By contrast, in the massive celestial objects, the MSW mechanism would be affected by the gravitational field and have
a different behavior. In this work, we are going to investigate the modification caused by the curved spacetime.

Various works have been dedicated to neutrino oscillations in curved spacetime; it was first pioneered by Stodolsky\cite{stodolsky79} and then discussed in detail by many authors\cite{ahluwalia96,piriz96,fornengo97,cardall97,ahluwalia97,konno98,habib99,fornengo99,wudka01,ahluwalia04,maiwa04,lambiase05,geralico12,visinelli14}. There are several methods developed and applied: plane wave method\cite{fornengo97,konno98,fornengo99}, WKB approximation\cite{stodolsky79,wudka01,maiwa04,visinelli14} and geometric treatment\cite{cardall97}. For various metrics: such as Schwarzschild metric\cite{ahluwalia96,piriz96,fornengo97,cardall97,ahluwalia97,habib99,fornengo99}, Kerr metric\cite{konno98,wudka01}, Kerr-Newman metric\cite{visinelli14}, Hartle-Thorne metric\cite{geralico12}, Lense-Thirring metric\cite{lambiase05}, and for different celestial objects: active galactic nuclei\cite{piriz96}, neutron stars\cite{ahluwalia96} and supernovae\cite{ahluwalia96,ahluwalia04},  discussions are made. The possible spin oscillation in gravitational field is investigated later\cite{dvornikov02,dvornikov05,dvornikov06,dvornikov13,alavi13}.

Most of the former literatures are focused on the two-generation oscillations  and to the best of our knowledge, three-generation oscillations and the MSW effect and the corresponding neutrino self-energy correction in curved spacetime have not been studied yet.  This topic is directly linked to one of the most fundamental yet undetermined features of neutrinos; {\it Mass Hierarchy Problem}. One of the two major experimental approaches\cite{qian15} to solve this problem is observing the MSW effect
which would clearly indicate how the conversion probability is related to the sign of $\mathrm{\Delta}{m}_{\mathrm{31}}^{2}$. Furthermore, determining this puzzle may play a significant role for understanding the nature of neutrinos (Dirac or Marjonara). In this paper we present a preliminary discussion on the topic.

In Sec.(\ref{s2}) we give a short review of the geometric treatment of the phase differences in neutrino oscillations. In Sec.(\ref{s3}) we discuss the possible corrections of the neutrino self-energy in curved spacetime briefly. In Sec.(\ref{s4}) and Sec.(\ref{s5}) we make explicit calculations on three-generation neutrino oscillations in vacuum and matter in a static, spherically symmetric spacetime.
In Sec.(\ref{s6}) the numerical estimations of the strength of such modifications are carried out and our conclusion is summarized in Sec.(\ref{s7}). It is noted that here we choose the sign of metric to be $diag(-,+,+,+)$ and use the natural unit system with $\hbar=c=1$.

\section{Geometric effects on the phase differences}\label{s2}
Two critical points concerning neutrino oscillations in curved
spacetime are the definitions of the energy and phase difference
occurring in the formulation of neutrino oscillations. Here we apply
a framework for curved spacetime where the geometric effect
manifests\cite{cardall97}, the energy is defined clearly
and the phase difference is covariantly expressed. For the convenience
of readers, only results of this treatment are listed here. The concerned formalism and notations are reviewed in
the appendix.

The original expression of the one-particle-neutrino wave function in
flat spacetime is
\begin{gather}
\left.{\left|{{\mathrm{\Psi}}_{\mathit{\alpha}}\mathrm{(}x\mathrm{,}t\mathrm{)}}\right.}\right\rangle\mathrm{{=}}\mathop{\sum}\limits_{j}{{U}_{\mathit{\alpha}{j}}}{e}^{i{P}^{\mathit{\mu}}{x}_{\mathit{\mu}}}\left.{\left|{{\mathit{\nu}}_{j}}\right.}\right\rangle,\label{originE}
\end{gather}
where
$\left.{\left|{{\mathrm{\Psi}}_{\mathit{\alpha}}\mathrm{(}x\mathrm{,}t\mathrm{)}}\right.}\right\rangle$
denotes the wave function in the flavor representation and is
expanded into
$\left.{\left|{{\mathrm{\Psi}}_{\mathit{\alpha}}\mathrm{(}x\mathrm{,}t\mathrm{)}}\right.}
\right\rangle\mathrm{{=}}\mathop{\sum}\limits_{j}{{U}_{\mathit{\alpha}{j}}}\left.{\left|{{\mathit{\nu}}_{j}}\right.}\right\rangle$ with $\left.{\left|{{\mathit{\nu}}_{j}}\right.}\right\rangle$ being the wave function in the
mass representation. The roman letters $\mathit{\alpha}\mathrm{{=}}\mathit{\nu}{\mathrm{,}}\mathit{\mu}{\mathrm{,}}\mathit{\tau}$ and the latin letters ${j}\mathrm{{=}}{1}{\mathrm{,}}{2}{\mathrm{,}}{3}$ denote flavor and mass states, respectively. ${U}_{\mathit{\alpha}{j}}$ is the element of the neutrino mixing matrix (PMNS matrix). ${P}^{\mathit{\mu}}$ is the four-momentum of neutrino in the mass basis and ${e}^{i{P}^{\mathit{\mu}}{x}_{\mathit{\mu}}}$ is the vacuum evolution phase which reduces to $e^{iEt}$ in the non-relativistic Schr\"odinger framework.

Generalizing Eq.(\ref{originE}) for the environment filled with electrons in a gravitational field, the formalism would be modified to be an equation in the curved spacetime as
\begin{gather}
\left.{\left|{{\mathrm{\Psi}}_{\mathit{\alpha}}{\mathrm{(}}\mathit{\tau}{\mathrm{)}}}\right.}\right\rangle\mathrm{{=}}\mathop{\sum}\limits_{j}{{U}_{\mathit{\alpha}{j}}{e}^{i\int{{\mathrm{(}}\frac{{M}^{2}}{2}\mathrm{{+}}{p}^{\mathit{\mu}}{A}_{\mathit{\mu}}{\mathrm{)}}{d}\mathit{\tau}}}}\left.{\left|{{\mathit{\nu}}_{j}}\right.}\right\rangle,\label{curvedE}
\end{gather}
where the position and time coordinates are replaced by an affine parameter $\tau$, ${M}^{2}$ is an operator acting on neutrino wave functions and its eigenvalues are the square of neutrino masses. ${p}^{\mathit{\mu}}$ is the tangent vector to the path along which neutrinos travel. This tangent vector is chosen so that ${p}^{0}\mathrm{{=}}{P}^{0}$ and the rest three spatial components are parallel to the three spatial components of $P^{\mu}$. ${A}_{\mathit{\mu}}$ is the effective potential caused by the electron environment. We can define the evolution phases of the corresponding wave functions as
\begin{gather}
{\mathrm{\Omega}}\mathrm{{\equiv}}{e}^{i\int{{\mathrm{(}}\frac{{M}^{2}}{2}\mathrm{{+}}{p}^{\mathit{\mu}}{A}_{\mathit{\mu}}{\mathrm{)}}{d}\mathit{\tau}}}.
\end{gather}

Differentiating  Eq.(\ref{curvedE}) a Schr\"{o}dinger-like evolution equation is yielded
\begin{gather}
\mathrm{{-}}{i}\frac{\mathrm{\partial}\left.{\left|{{\mathrm{\Psi}}_{\mathit{\alpha}}{\mathrm{(}}\mathit{\tau}{\mathrm{)}}}\right.}\right\rangle}{\mathrm{\partial}\mathit{\tau}}\mathrm{{=}}{\mathrm{(}}\frac{{M}^{2}}{2}\mathrm{{+}}{p}^{\mathit{\mu}}{A}_{\mathit{\mu}}{\mathrm{)}}\left.{\left|{{\mathrm{\Psi}}_{\mathit{\alpha}}{\mathrm{(}}\mathit{\tau}{\mathrm{)}}}\right.}\right\rangle,\label{evolutionE}
\end{gather}
and it can be seen that treating $\frac{{M}^{2}}{2}\mathrm{{+}}{p}^{\mathit{\mu}}{A}_{\mathit{\mu}}$ as an effective Hamiltonian would be of convenience.

Then in the mass representation, using Eq.(\ref{curvedE}) one can write down the probability of conversion for a neutrino from $\alpha$ flavor to $\beta$ flavor after traveling a spacetime interval $\Delta s(\tau-\tau_0)$ as
\begin{gather}
{\mathcal{Q}}_{\mathit{\beta}\mathit{\alpha}}\mathrm{\equiv}{\left|{\left\langle{{\mathrm{\Psi}}_{\mathit{\beta}}}\right.{\mathrm{(}}\mathit{\tau}{\mathrm{)}}\left.{\left|{{\mathrm{\Psi}}_{\mathit{\alpha}}\mathrm{(}{\mathit{\tau}}_{0}\mathrm{)}}\right.}\right\rangle}\right|}^{2}\mathrm{{=}}\mathop{\sum}\limits_{i\mathrm{,}j}{{U}_{\mathit{\alpha}{i}}^{\mathrm{\dag}}}{U}_{\mathit{\beta}{i}}{U}_{\mathit{\alpha}{j}}{U}_{\mathit{\beta}{j}}^{\mathrm{\dag}}{e}^{i{\mathrm{\Omega}}_{ji}},
\end{gather}
where the phase difference ${\mathrm{\Omega}}_{ij}$ of two mass eigenstates is defined as
\begin{gather}
{\mathrm{\Omega}}_{ji}\mathrm{\equiv}\int{{\mathrm{(}}\frac{{m}_{j}^{2}\mathrm{{-}}{m}_{i}^{2}}{2}\mathrm{{+}}\Delta{p}^{\mathit{\mu}}{A}_{\mathit{\mu}}{\mathrm{)}}{d}\mathit{\tau}}\mathrm{\equiv}\int{{\mathrm{(}}\frac{{\mathrm{\Delta}}_{ji}}{2}\mathrm{{+}}\Delta{p}^{\mathit{\mu}}{A}_{\mathit{\mu}}{\mathrm{)}}{d}\mathit{\tau}},\label{pd}
\end{gather}
where ${\mathrm{\Delta}}_{ji}\mathrm{\equiv}{m}_{j}^{2}\mathrm{{-}}{m}_{i}^{2}$ denotes the mass difference and $m_j$ is the eigenvalue of the corresponding mass eigenstate. $\mathrm{\Delta}{p}^{\mathit{\mu}}{A}_{\mathit{\mu}}$ is the difference of the electron-environmental contribution due to the asymmetric matter effect, where W boson exchange only affects electron neutrinos. Noticing these quantities are expressed in the mass basis and so that ${M}^{2}\mathrm{{=}}{diag}{\mathrm{(}}{m}_{1}^{2}{\mathrm{,}}{m}_{2}^{2}{\mathrm{,}}{m}_{3}^{2}{\mathrm{)}}$, we are able to write down Eq.(\ref{pd}). This phase difference and mixing matrix will compose the sufficient condition for calculating the conversion probability.

\section{Neutrino self-energy in curved spacetime}\label{s3}
We now turn to the specific expression of the effective potential ${A}_{\mathit{\mu}}$ which is induced by the corrections to the neutrino self-energy in the electron environment. The leading order contribution to the self-energy arises from the charged current and neutral current interactions. The later one has exactly the same effect on all the three neutrino generations, thus shows no significance to the MSW effect and can be neglected.
\begin{figure}[htbp]
\centering
\includegraphics[width=0.5\textwidth]{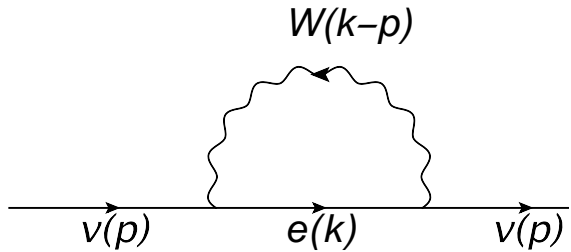}
\caption{Feynman diagram of charged current interaction of electron neutrinos}\label{fig1}
\end{figure}

As is shown in Fig.(\ref{fig1}), the Feynman diagram of the contribution from the charged current interaction (W boson exchange) is presented. To compute this diagram in curved spacetime, two modifications should be accounted.
\begin{enumerate}
\item {\it Finite-temperature field theory in curved spacetime}\cite{hu82,hu83,hu87,landsman86}. In the gravitational background, the definitions of both vacua and thermal equilibria become ambiguous; a vacuum state in equilibrium seen by one observer may appear as a state with particles and deviated from equilibrium\cite{semenoff85}. Besides, the possible phase transitions are also of interest. To solve these difficulties a finite-temperature field theory in curved spacetime is needed.
\item {\it Quantum field theory in curved spacetime and particle creation}\cite{parker68,hawking75}. As mentioned above, the definition of vacuum is not consistent with diffeomorphisms. Since particles are created via the coupling between gravitational field and quantum field, the Fock space on which the field operators act is changed, thus one may have to distinguish between the {\it in}-state and {\it out}-state Fock space. To avoid that ambiguity, one has to construct two sets of field operators and vacuum states corresponding to the {\it in} and the {\it out} states, respectively. The unique method for defining such states is still absent at present. Furthermore, due to these {\it in} and {\it out} states there exist two types of the S-matrix elements in the amplitudes. Thus the Feynman rules for accounting the self-energy of neutrinos in the electroweak theory should be modified.
\end{enumerate}

These two effects are most important for discussions in some cases such as the expanding universe, where the dynamic spacetime deviates the quantum states from equilibrium, or the ultra-strong gravitational field, where the particle creation becomes possible. Yet if we restrict ourselves to neutrino oscillations in celestial objects like sun or supernovae, the calculation can be substantially simplified. Besides, the region where the  MSW resonance appears can be very narrow\cite{smirnov03}, so the variation of the metric inside this region can be neglected. Thus it can be assumed that the environmental matter is always in equilibrium and no particle creation takes place in this case. Only propagators should be modified while vertices and Feynman rules remain the same.

Under these approximations and keeping the propagators in the momentum representation, we can write down the common Feynman rules for Fig.(\ref{fig1}) as
\begin{gather}
\mathrm{{-}}{i}\mathrm{\Sigma}{\mathrm{(}}{\mathbf{p}}{\mathrm{)}}\mathrm{{=}}\mathrm{{-}}{i}\int{{d}^{4}{k}{\mathrm{(}}\mathrm{{-}}\frac{i}{\sqrt{2}}{g}_{W}{\mathrm{)}}}{\mathit{\gamma}}^{\mathit{\mu}}{\mathcal{P}}_{L}{iS}_{e}{\mathrm{(}}{\mathbf{k}}{\mathrm{)(}}\mathrm{{-}}\frac{i}{\sqrt{2}}{g}_{W}{\mathrm{)}}{\mathit{\gamma}}^{\mathit{\mu}}{\mathcal{P}}_{L}\frac{i}{{\mathrm{(}}{\mathbf{k}}\mathrm{{-}}{\mathbf{p}}{\mathrm{)}}^{2}\mathrm{{-}}{m}_{W}^{2}},\label{se}
\end{gather}
where ${g}_{W}$ is the coupling constant of SU(2), ${\mathcal{P}_{L}}$ is the left-projection operator, $\gamma ^\mu$ is the gamma matrix in curved spacetime and is defined in Eq.(\ref{gamma}), ${iS}_{e}$ is the propagator of electrons in  thermal bath with gravitational background. The W boson propagator has been written in 't Hooft-Feynman gauge. With the gravitational modifications, the propagator of spin-$1/2$ fermions written in the momentum representation reads\cite{bunch79}
\begin{gather}
{i}{S}{\mathrm{(}}{\mathbf{k}}{\mathrm{)}}\mathrm{{=}}{\mathrm{(}}{1}\mathrm{{+}}{f}_{1}{\mathrm{(}}{x}^{\mathit{\mu}}{\mathrm{)(}}\mathrm{{-}}\frac{\mathrm{\partial}}{\mathrm{\partial}{m}^{2}}{\mathrm{)}}\mathrm{{+}}{f}_{2}{\mathrm{(}}{x}^{\mathit{\mu}}{\mathrm{)(}}\mathrm{{-}}\frac{\mathrm{\partial}}{\mathrm{\partial}{m}^{2}}{\mathrm{)}}^{2}{\mathrm{)}}{i}{S}_{0}{\mathrm{(}}{\mathbf k}{\mathrm{)}},\label{epcv}
\end{gather}
where $f_1$ and $f_2$ are functions of $x^\mu$, ${iS}_{0}\mathrm{(}k\mathrm{)}$ is the corresponding electron propagator in thermal bath with spacetime being flat. Considering modifications up to the order of $\frac{\mathrm{\partial}}{\mathrm{\partial}{m}^{2}}$ only
and ignoring the higher oder terms in $f_{1}(x^{\mu})$ which are dependent on the electron momentum,  the electron propagator in curved spacetime is
\begin{gather}
{iS}{\mathrm{(}}{\mathbf k}{\mathrm{)}}\mathrm{{=}}{\mathrm{(}}{1}\mathrm{{+}}\frac{1}{\mathrm{12}}{\mathcal{R}}{\mathrm{(}}\mathrm{{-}}\frac{\mathrm{\partial}}{\mathrm{\partial}{m}^{2}}{\mathrm{))}}{iS}_{0}{\mathrm{(}}{\mathbf k}{\mathrm{)}},
\end{gather}
where $\mathcal{R}$ is the Ricci scalar curvature of the spacetime.

Writing down the flat-spacetime propagator of electron field in terms of the real-time formalism with finite temperature\cite{dolan74}:
\begin{gather}
{iS}_{0}{\mathrm{(}}{\mathbf k}{\mathrm{)}}\mathrm{{=}}{\mathrm{(}}{\mathit{\gamma}}^{\mathit{\mu}}{k}_{\mathit{\mu}}\mathrm{{+}}{m}_{e}{\mathrm{)(}}\frac{i}{{k}^{2}\mathrm{{-}}{m}_{e}{}^{2}}\mathrm{{-}}{2}\mathit{\pi}\mathit{\delta}{\mathrm{(}}{k}^{2}\mathrm{{-}}{m}_{e}{}^{2}{\mathrm{)}}{f}_{F}{\mathrm{(}}{k}_{\mathit{\mu}}{u}^{\mathit{\mu}}{\mathrm{))}}\label{epftd},
\end{gather}
where the mass of fermion $m$ has been replaced by the mass of electron $m_e$ and ${f}_{F}(x)$ represents the Fermi distribution function
\begin{gather}
{f}_{F}{\mathrm{(}}{x}{\mathrm{)}}\mathrm{{=}}\frac{\mathit{\theta}{\mathrm{(}}{x}{\mathrm{)}}}{{e}^{\mathit{\beta}{\mathrm{(}}{x}\mathrm{{-}}\mathrm{\mu}{\mathrm{)}}}\mathrm{{+}}{1}}\mathrm{{+}}\frac{\mathit{\theta}{\mathrm{(}}\mathrm{{-}}{x}{\mathrm{)}}}{{e}^{\mathrm{{-}}\mathit{\beta}{\mathrm{(}}{x}\mathrm{{-}}\mathrm{\mu}{\mathrm{)}}}\mathrm{{+}}{1}},
\end{gather}
where $\mathit{\beta}$ is the inverse of temperature, $\mathrm{\mu}$ is the chemical potential of the electron and $\theta(x)$ is the Heaviside step function.

Noticing that in Eq.(\ref{epftd}), the first term is the common
propagator in vacuum except modified gamma matrices being introduced
and  it may lead to infinity when the loop diagrams involving the
propagator are integrated. The renormalization schemes in curved
spacetime have been discussed in some details by several authors
\cite{bunch79,panangaden81}. Under the condition that temperature is
just above the threshold of thermal effects on electron, the propagator of W boson is unchanged and the
dispersion relation of electron is non-relativistic
${\mathit{\omega}}_{k}\mathrm{{=}}\tfrac{{k}^{2}}{2{m}_{e}}$ where
${\mathit{\omega}}_{k}$ is the energy of electron.

Then Eq.(\ref{se}) can be evaluated up to the order of $1\mathrm{/}{m}_{W}^{2}$ as
\begin{gather}
\mathrm{\Sigma}{\mathrm{(}}{\mathbf{p}}{\mathrm{)}}\mathrm{{=}}{\mathrm{(}}{1}\mathrm{{-}}\frac{\mathcal{R}}{\mathrm{16}{m}_{e}^{2}}{\mathrm{)}}\sqrt{2}{G}_{F}{N}_{e}{e}_{a}^{\mathit{\mu}}{u}^{a},
\label{selfenergy}
\end{gather}
where $G_F$ is the Fermi coupling constant, ${N}_{e}$ is the locally measured electron number density and ${u}^{a}$ is the four-velocity of the environmental electron current.  Comparing with the common result obtained in flat spacetime, the self-energy is modified by the tetrad ${e}_{a}^{\mathit{\mu}}$ and an extra term $1\mathrm{{-}}\frac{R}{\mathrm{16}{m}_{e}^{2}}$ arising from gravitation background. This term can be separated as an extra gravitational phase.

\section{Neutrino propagation in vacuum with curved spacetime}\label{s4}
In this section we consider the neutrino propagation in a general static and spherically symmetric spacetime, the metric of which is\cite{hobson06}
\begin{gather}
{ds}^{2}\mathrm{{=}}\mathrm{{-}}{e}^{{2}\mathrm{\Phi}}{dt}^{2}\mathrm{{+}}{e}^{{2}\mathrm{\Lambda}}{dr}^{2}\mathrm{{+}}{r}^{2}{d}{\mathit{\theta}}^{2}\mathrm{{+}}{r}^{2}{\sin}^{2}\mathit{\theta}{d}{\mathit{\phi}}^{2},\label{metric}
\end{gather}
where $\Phi(r)$ and $\Lambda(r)$ are arbitrary functions of the radial coordinate $r$. In such a spherically symmetric spacetime the contribution of spin connection vanishes\cite{cardall97,wheeler57}. Thus we can safely concentrate on flavor oscillations of neutrino and ignore possible spin flips. Another characteristic of this spacetime is the existence of a Killing vector ${\mathrm{\partial}}_{t}$ which corresponds to a conserved quantity ${P}_{0}\mathrm{{=}}{g}_{{0}\mathit{\upsilon}}{P}^{\mathit{\upsilon}}\mathrm{\equiv}\mathrm{{-}}{E}_{\mathrm{\infty}}$. As will be shown later, this quantity is just the energy of neutrino measured by an observer at ${r}\mathrm{{=}}\mathrm{\infty}$.

Without losing generality, here we consider only the radial propagation of neutrino in vacuum with curved spacetime, the integral element for the propagating phase  is written as a function of proper length Eq.(\ref{geode2}):
\begin{gather}
\begin{split}
{d}\mathit{\tau}\mathrm{{=}}&{dl}{\mathrm{(}}\mathrm{{-}}{g}_{\mathrm{00}}{\mathrm{(}}\frac{{dx}^{0}}{{d}\mathit{\tau}}{\mathrm{)}}^{2}{\mathrm{)}}^{\mathrm{{-}}{1}{\mathrm{/}}{2}}\\ \mathrm{{=}}&{dl}\frac{1}{{E}_{\mathrm{\infty}}{e}^{\mathrm{{-}}\mathrm{\Phi}}},
\end{split}
\end{gather}
where ${E}_{\mathrm{\infty}}{e}^{\mathrm{{-}}\mathrm{\Phi}}$ is understood as the local energy which is measured by an observer in the local inertial frame. $
{e}^{\mathrm{{-}}\mathrm{\Phi}}
$
vanishes as ${r}\mathrm{\rightarrow}\mathrm{\infty}$ , leaving $
{E}_{\mathrm{\infty}}
$ only, which means it is the energy measured at $r=\infty$
.The proper length Eq.(\ref{pl}) for radial propagation case reads
\begin{gather}
{dl}\mathrm{{=}}\sqrt{{g}_{\mathrm{11}}\mathrm{(}{dx}^{1}{\mathrm{)}}^{2}}.
\end{gather}

Then in vacuum without the electron environment, the phase difference Eq.(\ref{pd}) is calculated as
\begin{gather}
\begin{split}
{\mathrm{\Omega}}_{ji}\mathrm{{=}}&\int{\frac{{\mathrm{\Delta}}_{ji}}{2}{d}\mathit{\tau}}\\ \mathrm{{=}}&\int{\frac{{\mathrm{\Delta}}_{ji}}{{E}_{\mathrm{\infty}}{e}^{\mathrm{{-}}\mathrm{\Phi}}}{e}^{\mathrm{\Lambda}}dr}.
\end{split}
\end{gather}

This phase difference is exactly the same as which in the two-generation case except the labels {\it i,j} range from 1 to 3 now.

\section{Three-generation MSW effect in curved spacetime}\label{s5}
Using the metric Eq.({\ref{metric}}), assuming the electron
environment is at rest with respect to the oscillation process and
choosing the tetrad to be $
{e}_{a}^{\mathit{\mu}}\mathrm{{=}}{diag}{\mathrm{(}}{e}^{\mathrm{{-}}\mathrm{\Phi}}
{\mathrm{,}}{e}^{\mathrm{{-}}\mathrm{\Lambda}}{\mathrm{,}}\frac{1}{r}{\mathrm{,}}\frac{1}{{r}\sin\mathit{\theta}}{\mathrm{)}}
$, the self-energy Eq.(\ref{selfenergy}) of the neutrino in environment
with curved spacetime can be calculated. Then instead of
Eq.(\ref{selfenergy}) one can turn to the  effective potential
\begin{gather}
{A}_{\mathit{\mu}{f}}\mathrm{{=}}\mbox{ $\mathrm{\left({\begin{array}{ccc}{{}\mbox{ $\sqrt{2}$}{e}^{\mathrm{-\Phi}}{G}_{F}{N}_{e}}{\mathrm{(}}{1}\mathrm{{-}}\frac{\mathcal{R}}{\mathrm{16}{m}_{e}^{2}}{\mathrm{)}}
&{0}&{0}\\{0}&{0}&{0}\\{0}&{0}&{0}\end{array}}\right)}$},\label{effectivepotential}
\end{gather}
where the subscript $f$ refers to the flavor basis. It is noted that this effective potential is expanded in the flavor basis and diagonalized.

To calculate the phase differences, one could directly substitute Eq.(\ref{effectivepotential}) into Eq.(\ref{pd}), yet in general it is more convenient if we consider the effective potential as a modification to the masses of neutrinos and calculate the resultant mixing parameters in medium with the effective masses. Then neutrinos would behave exactly like they were in vacuum except possessing different masses and different mixings.

First writing the evolution relation Eq.(\ref{propagation}) for $\alpha$-flavor neutrino in the flavor basis:
\begin{gather}
\left.{\left|{{\mathrm{\Psi}}_{\mathit{\alpha}}\mathrm{(}r\mathrm{)}}\right.}\right\rangle\mathrm{{=}}{e}^{\mathrm{{-}}{i}\int{\frac{1}{2{E}_{\mathrm{\infty}}}{\mathrm{(}}{e}^{\mathrm{\Phi}}{M}_{f}^{2}\mathrm{{+}}{V}_{f}{\mathrm{)}}{e}^{\mathrm{\Lambda}}{d}{r}}}\left.{\left|{{\mathrm{\Psi}}_{\mathit{\alpha}}\mathrm{(}{r}_{0}\mathrm{)}}\right.}\right\rangle,
\end{gather}
where
\begin{gather}
{M}_{f}^{2}\mathrm{{=}}{U}{M}_{m}^{2}{U}^{\mathrm{\dag}}
\end{gather}
is the mass matrix in the flavor representation, which is transformed from the mass-basis mass matrix by the PMNS matrix. The mass matrix only appears in the phase part of the neutrino evolution, thus only the relative differences of the eigenvalues matter. Therefore subtracting $m_1^2$ from the mass-basis mass matrix and using the mass differences $\Delta_{21},\Delta_{31}$ as new eigenvalues instead of $m_1^2,m_2^2,m_3^2$, we have
\begin{gather}
{M}_{m}^{2}\mathrm{{=}}\mbox{$\mathrm{\left({\begin{array}{ccc}{0}&{0}&{0}\\{0}&{{\mathrm{\Delta}}_{21}}&{0}\\{0}&{0}&{{\mathrm{\Delta}}_{31}}\end{array}}\right)}$}.\label{mmr}
\end{gather}

The PMNS matrix is parameterized with mixing angles $\theta_{12},\theta_{13},\theta_{23}$ and a CP violating phase $\delta$
\begin{gather}
\begin{split}
{U}\mathrm{{=}}&\mbox{ $\mathrm{\left({\begin{array}{ccc}{1}&{0}&{0}\\{0}&{{c}_{23}}&{{s}_{23}}\\{0}&{{-}{s}_{23}}&{{c}_{23}}\end{array}}\right)}$}\mbox{ $\mathrm{\left({\begin{array}{ccc}{{c}_{13}}&{0}&{{e}^{{-}{i}\mathit{\delta}}{s}_{13}}\\{0}&{1}&{0}\\{{-}{e}^{{-}{i}\mathit{\delta}}{s}_{13}}&{0}&{{c}_{13}}\end{array}}\right)}$}\mbox{ $\mathrm{\left({\begin{array}{ccc}{{c}_{12}}&{{s}_{12}}&{0}\\{{-}{s}_{12}}&{{c}_{12}}&{0}\\{0}&{0}&{1}\end{array}}\right)}$}\\
\mathrm{{=}}&\mbox{ $\mathrm{\left({\begin{array}{ccc}{{c}_{12}{c}_{13}}&{{c}_{13}{s}_{12}}&{{e}^{{-}{i}\mathit{\delta}}{s}_{13}}\\{{-}{c}_{23}{s}_{12}{-}{e}^{{i}\mathit{\delta}}{c}_{12}{s}_{13}{s}_{23}}&{{c}_{12}{c}_{23}{-}{e}^{{i}\mathit{\delta}}{s}_{12}{s}_{13}{s}_{23}}&{{c}_{13}{s}_{23}}\\{{-}{e}^{{i}\mathit{\delta}}{c}_{12}{c}_{23}{s}_{13}{+}{s}_{12}{s}_{23}}&{{-}{e}^{{i}\mathit{\delta}}{c}_{23}{s}_{12}{s}_{13}{-}{c}_{12}{s}_{23}}&{{c}_{13}{c}_{23}}\end{array}}\right)}$},
\label{para}
\end{split}
\end{gather}
where ${s}_{ij}\mathrm{\equiv}\sin{\mathit{\theta}}_{ij}$, ${c}_{ij}\mathrm{\equiv}\cos{\mathit{\theta}}_{ij}$. We set $\delta=0$ through this work.

And the electron-environmental contribution matrix $V_f$ is written as
\begin{gather}
{V}_{f}\mathrm{{=}}\mbox{ $\mathrm{\left({\begin{array}{ccc}{v(r)}&{0}&{0}\\{0}&{0}&{0}\\{0}&{0}&{0}\end{array}}\right)}$},
\end{gather}
where ${v}{\mathrm{(}}{r}{\mathrm{)}}\mathrm{{=}}{2}\sqrt{2}{e}^{-\Phi}{G}_{F}{E}_{\mathrm{\infty}}{N}_{e}{\mathrm{(}}{1}\mathrm{{-}}\frac{\mathcal{R}}{\mathrm{16}{m}_{e}^{2}}{\mathrm{)}}$.

Defining the effective mass matrix as
\begin{gather}
{\tilde{M}}_{f}^{2}\mathrm{{=}}{\mathrm{(}}{e}^{\mathrm{\Phi}}{M}_{f}^{2}\mathrm{{+}}{V}_{f}{\mathrm{)}}{e}^{\mathrm{\Lambda}},\label{mtilde}
\end{gather}
we obtain the three eigenvalues of the effective mass matrix
\begin{gather}
{x}_{A}\mathrm{{=}}{2}\sqrt{\frac{l}{3}}{\mathrm{cos}}{\mathrm{(}}\frac{1}{3}\mathrm{arccos}{\mathrm{(}}\frac{3s}{2l}\sqrt{\frac{l}{3}}{\mathrm{))}}\mathrm{{-}}\frac{b}{3a},\\
{x}_{B}\mathrm{{=}}{2}\sqrt{\frac{l}{3}}{\mathrm{cos}}{\mathrm{(}}\frac{1}{3}\mathrm{arccos}{\mathrm{(}}\frac{3s}{2l}\sqrt{\frac{l}{3}}{\mathrm{)}}\mathrm{{-}}\frac{{2}\mathit{\pi}}{3}{\mathrm{)}}\mathrm{{-}}\frac{b}{3a},\\
{x}_{C}\mathrm{{=}}{2}\sqrt{\frac{l}{3}}{\mathrm{cos}}{\mathrm{(}}\frac{1}{3}\mathrm{arccos}{\mathrm{(}}\frac{3s}{2l}\sqrt{\frac{l}{3}}{\mathrm{)}}\mathrm{{-}}\frac{{4}\mathit{\pi}}{3}{\mathrm{)}}\mathrm{{-}}\frac{b}{3a},
\end{gather}
where the parameters $a, b, c, d, s, l$ are defined as
\begin{align}
{a}&\mathrm{{=}}{8},\\
{b}&\mathrm{{=}}\mathrm{{-}}{8}{v}\mathrm{{-}}{8}{e}^{\mathrm{\Psi}}{\mathrm{\Delta}}_{\mathrm{21}}\mathrm{{-}}{8}{e}^{\mathrm{\Phi}}{\mathrm{\Delta}}_{\mathrm{31}},
\end{align}
\begin{gather}
\begin{split}
{c}\mathrm{{=}}&{e}^{\mathrm{\Phi}}{\mathrm{(}}{6}{v}{\mathrm{\Delta}}_{\mathrm{21}}\mathrm{{+}}{4}{v}{\mathrm{\Delta}}_{\mathrm{31}}\mathrm{{+}}{8}{e}^{\mathrm{\Phi}}{\mathrm{\Delta}}_{\mathrm{21}}{\mathrm{\Delta}}_{\mathrm{31}}\mathrm{{+}}{2}{v}{\mathrm{\Delta}}_{\mathrm{21}}\cos{2}{\mathit{\theta}}_{\mathrm{12}}\mathrm{{+}}\\
&{v}{\mathrm{\Delta}}_{\mathrm{21}}\cos{2}\mbox{ $\mathrm{\left({{\mathit{\theta}}_{12}{-}{\mathit{\theta}}_{13}}\right)}$}\mathrm{{-}}{2}{v}{\mathrm{\Delta}}_{\mathrm{21}}\cos{2}{\mathit{\theta}}_{\mathrm{13}}\mathrm{{+}}{4}{v}{\mathrm{\Delta}}_{\mathrm{31}}\cos{2}{\mathit{\theta}}_{\mathrm{13}}\mathrm{{+}}{v}{\mathrm{\Delta}}_{\mathrm{21}}\cos{2}\mbox{ $\mathrm{\left({{\mathit{\theta}}_{12}{+}{\mathit{\theta}}_{13}}\right)}$}{\mathrm{)}},
\end{split}
\end{gather}
\begin{align}
{d}& \mathrm{{=}}\mathrm{{-}}{8}{e}^{{2}\mathrm{\Phi}}{v}{\mathrm{\Delta}}_{\mathrm{21}}{\mathrm{\Delta}}_{\mathrm{31}}\cos{\mathit{\theta}}_{\mathrm{12}}{}^{2}\cos{\mathit{\theta}}_{\mathrm{13}}{}^{2},\\
{s}& \mathrm{{=}}\mathrm{\frac{{2}{b}^{3}{-}{9}{abc}{+}{27}{a}^{2}{d}}{27{a}^{3}}},\\
{l}& \mathrm{{=}}\mathrm{{-}}\frac{{3}{ac}\mathrm{{-}}{b}^{2}}{3{a}^{2}}.
\end{align}

It should be clarified that ${x}_{A}\mathrm{,}{x}_{B}\mathrm{,}{x}_{C}$ are the three eigenvalues of matrix (\ref{mtilde}) with the order ${x}_{A}\mathrm{\geq}{x}_{B}\mathrm{\geq}{x}_{C}$. Thus they may not be in a sequence of ${m}_{1}\mathrm{,}{m}_{2}\mathrm{,}{m}_{3}$, nor do they have the same values of the real neutrino masses due to the simplification of matrix (\ref{mmr}). Only the values of their differences are identical to the corresponding mass differences. Because of this ambiguity, we need to discuss these solutions for normal hierarchy (NH) and inverted hierarchy (IH) respectively.

For the normal hierarchy, there are
\begin{gather}
{\tilde{m}}_{1}^{2}\mathrm{{=}}{x}_{C}\mathrm{{+}}{C}{\mathrm{,}}{\tilde{m}}_{2}^{2}\mathrm{{=}}{x}_{B}\mathrm{{+}}{C}{\mathrm{,}}{\tilde{m}}_{3}^{2}\mathrm{{=}}{x}_{A}\mathrm{{+}}{C},
\end{gather}
where ${C}\mathrm{{=}}{m}_{1}^{2}$ is caused by the subtraction of mass ${m}_{1}^{2}$ from the mass matrix.

And so for the inverted hierarchy, we have
\begin{gather}
{\tilde{m}}_{3}^{2}\mathrm{{=}}{x}_{C}\mathrm{{+}}{C}{\mathrm{,}}{\tilde{m}}_{1}^{2}\mathrm{{=}}{x}_{B}\mathrm{{+}}{C}{\mathrm{,}}{\tilde{m}}_{2}^{2}\mathrm{{=}}{x}_{A}\mathrm{{+}}{C}.
\end{gather}


Solving the resonance condition for $v$ yields
 \begin{gather}
{\left.{v}\right|}_{res}\mathrm{{=}}{e}^{\mathrm{\Phi}}{\left.{{v}_{flat}}\right|}_{res},
\end{gather}
where ${\left.{{v}_{flat}}\right|}_{res}$ denotes the resonance condition in the Minkowski spacetime with the neutrino energy being ${\left.{{E}_{flat}}\right|}_{res}$. Substituting the expression of $v$, it is obtained that
\begin{gather}
{\left.{{E}_{\mathrm{\infty}}}\right|}_{res}\mathrm{{=}}{e}^{{2}\mathrm{\Phi}}{\mathrm{(}}{1}\mathrm{{-}}\frac{R}{\mathrm{16}{m}_{e}^{2}}{\mathrm{)}}^{\mathrm{{-}}{1}}{\left.{{E}_{flat}}\right|}_{res}.
\end{gather}

The modification to the resonance energy consists of two parts, the overall redshift with a
factor ${e}^{{2}\mathrm{\Phi}}$ and an extra term coming from the gravitational effect on the neutrino self-energy.
The overall redshift factor ${e}^{{2}\mathrm{\Phi}}$  obtained here is the square of the redshift factor given in \cite{cardall97}.
\subsection{Neutrino mixing parameters in celestial objects}
It would be convenient to write the medium-modified mixing matrix in the same parametrization scheme as Eq.(\ref{para}), then the oscillation in matter will be of the same form as the oscillation in vacuum.

We can define the effective phase and phase difference in medium with results obtained above to be
\begin{gather}
{\tilde{\mathrm{\Omega}}}_{j}\mathrm{\equiv}\int{\frac{{\tilde{m}}_{j}^{2}}{{E}_{\mathrm{\infty}}{e}^{\mathrm{{-}}\mathrm{\Phi}}}{e}^{\mathrm{\Lambda}}dr},\\
{\tilde{\mathrm{\Omega}}}_{ji}\mathrm{{\equiv}}\int{\frac{{\tilde{\mathrm{\Delta}}}_{ji}}{{E}_{\mathrm{\infty}}{e}^{\mathrm{{-}}\mathrm{\Phi}}}{e}^{\mathrm{\Lambda}}dr}.
\end{gather}

To calculate the mixing matrix in medium one needs the time-dependent perturbation theory and the evolution equation Eq.(\ref{evolutionE})
\begin{gather}
\mathrm{{-}}{i}\frac{\mathrm{\partial}}{\mathrm{\partial}\mathit{\tau}}{\tilde{U}}e^{i{\mathrm{\Omega}}_{j}}\left.{\left|{{\mathit{\nu}}_{j}}\right.}\right\rangle\mathrm{{=}}{\tilde{M}}_{m}^{2}{\tilde{U}}e^{i{\mathrm{\Omega}}_{j}}\left.{\left|{{\mathit{\nu}}_{j}}\right.}\right\rangle,\label{evolutionEM}
\end{gather}
where ${\tilde{\mathrm{\Delta}}}_{ji}\mathrm{\equiv}{\tilde{m}}_{j}^{2}\mathrm{{-}}{\tilde{m}}_{i}^{2}$ is the effective mass difference in medium.
The elements of the effective mixing matrix obey differential equations
\begin{gather}
\mathrm{{-}}{i}\frac{\mathrm{\partial}}{\mathrm{\partial}\mathit{\tau}}\mathop{\sum}\limits_{j}{{\tilde{U}}_{\mathit{\alpha}{j}}}\mathrm{{=}}\mathop{\sum}\limits_{j}{{V}_{mij}{\tilde{U}}_{\mathit{\alpha}{j}}{e}^{i{\tilde{\mathrm{\Omega}}}_{ji}}}.\label{mixingE}
\end{gather}

In general it is difficult to solve for these mixing parameters analytically and in practice one would do it numerically. But if the envoronmental matter has uniform density and the first order derivative of metric can be neglected, Eq.(\ref{evolutionEM}) can be rewritten as
\begin{gather}
\mathrm{{-}}{i}\frac{\mathrm{\partial}}{\mathrm{\partial}\mathit{\tau}}\mathop{\sum}\limits_{j}{{\tilde{U}}_{\mathit{\alpha}{j}}}{e}^{i{\tilde{\mathrm{\Omega}}}_{j}}\left.{\left|{{\mathit{\nu}}_{j}}\right.}\right\rangle\mathrm{{=}}{\tilde{U}}^{\mathrm{\dag}}{\tilde{M}}_{m}^{2}\tilde{U}\mathop{\sum}\limits_{j}{{e}^{i{\tilde{\mathrm{\Omega}}}_{j}}\left.{\left|{{\mathit{\nu}}_{j}}\right.}\right\rangle}.\label{ce1}
\end{gather}
Then the modified PMNS matrix  reduces to be the unitary diagonalization matrix for ({\ref{mtilde}}) through the similarity relation
\begin{gather}
{\tilde{U}}^{\mathrm{\dag}}{\tilde{M}}_{f}^{2}\tilde{U}\mathrm{{=}}{\tilde{M}}_{m}^{2}.
\end{gather}

Solving for the diagonalization matrix through the Gram-Schmidt process then applying re-parameterization, the modified mixing angles can be calculated in either NH or IH. For NH, there are
\begin{gather}
{{}^{N}{\tilde{t}}_{\mathrm{12}}^{2}\mathrm{{=}}\mathrm{\frac{{\mbox{ $\left({{\mathit{\rho}}_{A}^{2}{\mathit{\xi}}_{B}{-}{\mathit{\xi}}_{A}{\mathit{\rho}}_{A}{\mathit{\rho}}_{B}{+}{\mathit{\sigma}}_{A}\left({{\mathit{\sigma}}_{A}{\mathit{\xi}}_{B}{-}{\mathit{\xi}}_{A}{\mathit{\sigma}}_{B}}\right)}\right)$}}^{2}}{\mbox{ $\left({{\mathit{\xi}}_{A}^{2}{+}{\mathit{\rho}}_{A}^{2}{+}{\mathit{\sigma}}_{A}^{2}}\right)$}{\mbox{ $\left({{\mathit{\sigma}}_{A}{\mathit{\rho}}_{B}{-}{\mathit{\rho}}_{A}{\mathit{\sigma}}_{B}}\right)$}}^{2}}}} \hfill,\\
{{}^{N}{\tilde{t}}_{\mathrm{13}}^{2}\mathrm{{=}}\mathrm{\frac{{\mathit{\xi}}_{A}^{2}}{{\mathit{\xi}}_{A}^{2}{+}{\mathit{\rho}}_{A}^{2}{+}{\mathit{\sigma}}_{A}^{2}}}} \hfill,\\
{{}^{N}{\tilde{t}}_{\mathrm{23}}^{2}\mathrm{{=}}\frac{{\mathit{\rho}}_{A}^{2}}{{\mathit{\sigma}}_{A}^{2}}} \hfill,
\end{gather}
where ${\tilde{t}}_{ij}^{2}\mathrm{\equiv}{\tan}^{2}{\tilde{\mathit{\theta}}}_{ij}$ and N denotes normal hierarchy. Similarly, for IH it yields

\begin{gather}
{{}^{I}{\tilde{t}}_{\mathrm{12}}^{2}\mathrm{{=}}\mathrm{\frac{{\mathit{\xi}}_{A}^{2}\mbox{ $\left({{\mathit{\sigma}}_{A}^{2}\left({{\mathit{\xi}}_{B}^{2}{+}{\mathit{\rho}}_{B}^{2}}\right){-}{2}{\mathit{\xi}}_{A}{\mathit{\sigma}}_{A}{\mathit{\xi}}_{B}{\mathit{\sigma}}_{B}{-}{2}{\mathit{\rho}}_{A}{\mathit{\rho}}_{B}\left({{\mathit{\xi}}_{A}{\mathit{\xi}}_{B}{+}{\mathit{\sigma}}_{A}{\mathit{\sigma}}_{B}}\right){+}{\mathit{\rho}}_{A}^{2}\left({{\mathit{\xi}}_{B}^{2}{+}{\mathit{\sigma}}_{B}^{2}}\right){+}{\mathit{\xi}}_{A}^{2}\left({{\mathit{\rho}}_{B}^{2}{+}{\mathit{\sigma}}_{B}^{2}}\right)}\right)$}}{{\mbox{ $\left({{\mathit{\rho}}_{A}^{2}{\mathit{\xi}}_{B}{-}{\mathit{\xi}}_{A}{\mathit{\rho}}_{A}{\mathit{\rho}}_{B}{+}{\mathit{\sigma}}_{A}\left({{\mathit{\sigma}}_{A}{\mathit{\xi}}_{B}{-}{\mathit{\xi}}_{A}{\mathit{\sigma}}_{B}}\right)}\right)$}}^{2}}}} \hfill,\\
{{}^{I}{\tilde{t}}_{\mathrm{13}}^{2}\mathrm{{=}}\mathrm{\frac{{\mbox{ $\left({{\mathit{\sigma}}_{A}{\mathit{\rho}}_{B}{-}{\mathit{\rho}}_{A}{\mathit{\sigma}}_{B}}\right)$}}^{2}}{{\mathit{\rho}}_{A}^{2}{\mathit{\xi}}_{A}^{2}{-}{2}{\mathit{\xi}}_{A}{\mathit{\rho}}_{A}{\mathit{\xi}}_{B}{\mathit{\rho}}_{B}{+}{\mathit{\xi}}_{A}^{2}{\mathit{\rho}}_{B}^{2}{+}{\mbox{ $\left({{\mathit{\sigma}}_{A}{\mathit{\xi}}_{B}{-}{\mathit{\xi}}_{A}{\mathit{\sigma}}_{B}}\right)$}}^{2}}}} \hfill,\\
{{}^{I}{\tilde{t}}_{\mathrm{23}}^{2}\mathrm{{=}}{\mathrm{(}}\mathrm{\frac{{\mathit{\sigma}}_{A}{\mathit{\xi}}_{B}{-}{\mathit{\xi}}_{A}{\mathit{\sigma}}_{B}}{{-}{\mathit{\rho}}_{A}{\mathit{\xi}}_{B}{+}{\mathit{\xi}}_{A}{\mathit{\rho}}_{B}}}{\mathrm{)}}^{2}} \hfill.
\end{gather}

The parameters appearing in the above equations are defined as
\begin{gather}
{\mathit{\xi}}_{I}\mathrm{\equiv}\mathrm{{-}}{\mathit{\zeta}}^{2}\mathrm{{+}}{\mathrm{(}}\mathrm{\varepsilon}\mathrm{{-}}{x}_{I}{\mathrm{)(}}\mathit{\eta}\mathrm{{-}}{x}_{I}{\mathrm{)}};\\
{\mathit{\rho}}_{I}\mathrm{\equiv}\mathit{\gamma}\mathit{\zeta}\mathrm{{+}}\mathit{\beta}{\mathrm{(}}{x}_{I}\mathrm{{-}}\mathit{\eta}{\mathrm{)}};\\
{\mathit{\sigma}}_{I}\mathrm{\equiv}\mathit{\beta}\mathit{\zeta}\mathrm{{+}}\mathit{\gamma}{\mathrm{(}}{x}_{I}\mathrm{{-}}\mathrm{\varepsilon}{\mathrm{)}},
\end{gather}
where $I=A,B,C$, and
\begin{gather}
{\mathit{\alpha}\mathrm{\equiv}{e}^{\mathrm{\Lambda}}{\mathrm{(}}{v}\mathrm{{+}}{e}^{\mathrm{\Psi}}{\mathrm{\Delta}}_{\mathrm{21}}{c}_{\mathrm{13}}^{2}{s}_{\mathrm{12}}^{2}\mathrm{{+}}{e}^{\mathrm{\Psi}}{\mathrm{\Delta}}_{\mathrm{31}}{s}_{\mathrm{13}}^{2}{\mathrm{)}}} \hfill;\\
{\mathit{\beta}\mathrm{\equiv}{e}^{\mathrm{\Lambda}\mathrm{{+}}\mathrm{\Psi}}{\mathrm{(}}{\mathrm{\Delta}}_{\mathrm{21}}{\mathrm{(}}{s}_{\mathrm{12}}{c}_{\mathrm{12}}{c}_{\mathrm{13}}{c}_{\mathrm{23}}\mathrm{{-}}{s}_{\mathrm{12}}^{2}{s}_{\mathrm{13}}{s}_{\mathrm{23}}{c}_{\mathrm{13}}{\mathrm{)}}\mathrm{{+}}{\mathrm{\Delta}}_{\mathrm{31}}{s}_{\mathrm{13}}{s}_{\mathrm{23}}{c}_{\mathrm{13}}{\mathrm{)}}} \hfill;\\
{\mathit{\gamma}\mathrm{\equiv}{e}^{\mathrm{\Lambda}\mathrm{{+}}\mathrm{\Psi}}{\mathrm{(}}\mathrm{{-}}{\mathrm{\Delta}}_{\mathrm{21}}{\mathrm{(}}{s}_{\mathrm{12}}{s}_{\mathrm{23}}{c}_{\mathrm{12}}{c}_{\mathrm{13}}\mathrm{{+}}{s}_{\mathrm{12}}^{2}{s}_{\mathrm{13}}{c}_{\mathrm{13}}{c}_{\mathrm{23}}{\mathrm{)}}\mathrm{{+}}{\mathrm{\Delta}}_{\mathrm{31}}{s}_{\mathrm{13}}{c}_{\mathrm{23}}{c}_{\mathrm{13}}{\mathrm{)}}} \hfill;\\
{\mathrm{\varepsilon}\mathrm{\equiv}{e}^{\mathrm{\Lambda}\mathrm{{+}}\mathrm{\Psi}}{\mathrm{(}}{\mathrm{\Delta}}_{\mathrm{21}}{\mathrm{(}}{c}_{\mathrm{12}}{c}_{\mathrm{23}}\mathrm{{-}}{s}_{\mathrm{12}}{s}_{\mathrm{13}}{s}_{\mathrm{23}}{\mathrm{)}}^{2}\mathrm{{+}}{\mathrm{\Delta}}_{\mathrm{31}}{s}_{\mathrm{23}}^{2}{c}_{\mathrm{13}}^{2}{\mathrm{)}}} \hfill;\\
{\mathit{\zeta}\mathrm{\equiv}{e}^{\mathrm{\Lambda}\mathrm{{+}}\mathrm{\Psi}}{\mathrm{(}}\mathrm{{-}}{\mathrm{\Delta}}_{\mathrm{21}}{\mathrm{(}}{s}_{\mathrm{23}}{c}_{\mathrm{12}}\mathrm{{+}}{s}_{\mathrm{12}}{s}_{\mathrm{13}}{c}_{\mathrm{23}}{\mathrm{)(}}{c}_{\mathrm{12}}{c}_{\mathrm{23}}\mathrm{{-}}{s}_{\mathrm{12}}{s}_{\mathrm{13}}{s}_{\mathrm{23}}{\mathrm{)}}\mathrm{{+}}{\mathrm{\Delta}}_{\mathrm{31}}{s}_{\mathrm{23}}{c}_{\mathrm{13}}^{2}{c}_{\mathrm{23}}{\mathrm{)}}} \hfill;\\
{\mathit{\eta}\mathrm{\equiv}{e}^{\mathrm{\Lambda}\mathrm{{+}}\mathrm{\Psi}}{\mathrm{(}}{\mathrm{\Delta}}_{\mathrm{21}}{\mathrm{(}}{s}_{\mathrm{12}}{s}_{\mathrm{13}}{c}_{\mathrm{23}}\mathrm{{+}}{s}_{\mathrm{23}}{c}_{\mathrm{12}}{\mathrm{)}}^{2}\mathrm{{+}}{\mathrm{\Delta}}_{\mathrm{31}}{c}_{\mathrm{13}}^{2}{c}_{\mathrm{23}}^{2}{\mathrm{)}}} \hfill.
\end{gather}
\section{Numerical estimations}\label{s6}
The actual structure of real celestial objects can be rather complicated and  this section we present an explicit estimate for a toy model where the matter density is uniform. All quantities are written in the SI units for numerical computations. The observer is located at $r=\infty$.

Now let us estimate the matter effect when neutrinos travel inside a massive celestial object. We need an interior Schwarzschild solution to describe the background spacetime. These solutions of the Einstein field equations are usually impossible to be explicitly written in a simple form but are presented numerically.  For simplicity, here we apply the easiest interior Schwarzschild solution in which the celestial object is approximated as an isotropic, uncharged non-rotating perfect fluid. The metric can be evaluated using the Oppenheimer-Volkoff equation and the result is\cite{hobson06}
\begin{gather}
{ds}^{2}\mathrm{{=}}\mathrm{{-}}{\mathrm{(}}\frac{3}{2}{\mathrm{(}}{1}\mathrm{{-}}{2}\frac{GM}{R{c}^{2}}{\mathrm{)}}^{1\mathrm{/}2}\mathrm{{-}}\frac{1}{2}{\mathrm{(}}{1}\mathrm{{-}}{2}\frac{{GMr}^{2}}{{R}^{3}{c}^{2}}{\mathrm{)}}^{1\mathrm{/}2}{\mathrm{)}}^{2}c^2{dt}^{2}\mathrm{{+}}\frac{1}{{1}\mathrm{{-}}\frac{2{GMr}^{2}}{{R}^{3}{c}^{2}}}{dr}^{2}\mathrm{{+}}{r}^{2}{d}{\mathit{\theta}}^{2}\mathrm{{+}}{r}^{2}{\sin}^{2}\mathit{\theta}{d}{\mathit{\phi}}^{2},\label{interiorS}
\end{gather}
where $M$ is the mass of the celestial object through which the neutrino traverses and $R$ is the radius of the object, $r$ is the radial coordinate and $r\mathrm{<}R$. To have a noticeable effect on the oscillation, we consider a celestial object whose radius is $10^2$ times larger than that of the sun and is $10^6$ times heavier than the sun. In this configuration the celestial object would have the same density as the sun. $M=10^6{M}_{\mathrm{\odot}}\approx{2}\mathrm{\times}{\mathrm{10}}^{\mathrm{36}}$ kg, $R=10^2{R}_{\mathrm{\odot}}\approx{7}\mathrm{\times}{\mathrm{10}}^{10}$ km and $\mathit{\rho}={\mathit{\rho}}_{\mathrm{\odot}}\approx{\mathrm{10}}$ g/cm${^3}$ are the mass, radius and density of the celestial object, respectively.

For oscillations that take place deep inside the object, i.e. at the region where $r<<R$, we obtain that
\begin{gather}
{{e}^{{2}\mathrm{\Lambda}}\sim{1}} \hfill,\\
{{e}^{\mathrm{2\Phi}}\sim{\mathrm{0.94}}}\hfill.
\end{gather}

Recalling that $\mathit{\rho}={\mathrm{(}}{N}_{n}\mathrm{{+}}{N}_{p}{\mathrm{)}}{m}_{n}$ and ${Y}_{e}=\frac{{N}_{e}}{{N}_{n}\mathrm{{+}}{N}_{p}}$, where ${m}_{n}$ is the mass of a nucleon, ${N}_{n}$ and
${N}_{p}$ are the number densities of neutron and proton, respectively. ${Y}_{e}$ is the electron fraction and is approximated well as $0.5$. We can write down the environmental contribution function $v$ as
\begin{gather}
{v}{\mathrm{(}}{r}{\mathrm{)}}{=}{1}{\mathrm{.}}{\mathrm{518}}\mathrm{\times}{\mathrm{10}}^{\mathrm{{-}}{\mathrm{13}}}{e}^{\mathrm{{-}}\mathrm{\Phi}}{E}_{\mathrm{\infty}}\mathit{\rho}{Y}_{e}{\mathrm{(}}{1}\mathrm{{-}}\mathrm{\frac{\mathcal{R}{\hbar}^{2}}{16{m}_{e}^{2}{c}^{2}}}{\mathrm{)}} ,
\end{gather}
where the unit of $v$ is eV and the density is written in g/cm$^3$, the Ricci scalar has the unit of m$^{-2}$.
\begin{figure}[htbp]
\centering
\includegraphics[width=0.5\textwidth]{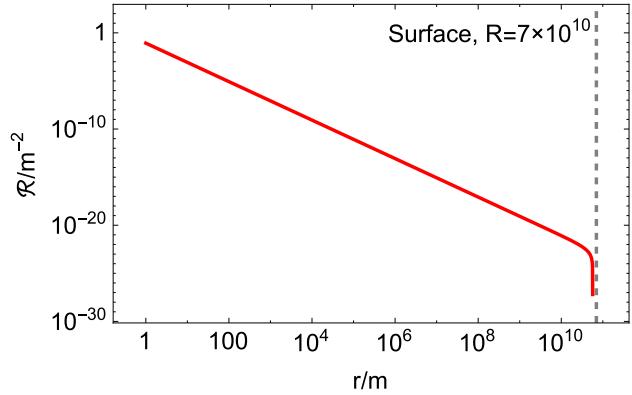}
\caption{Ricci scalar curvature vs. $r$}\label{fig2}
\end{figure}
For the metric (\ref{interiorS}) the Ricci scalar curvature is calculated as Fig.(\ref{fig2}) which has a nearly  exponential
dependence on $r$. It tends to infinity as $r\mathrm{\rightarrow}0$ and drops drastically as $r\mathrm{\rightarrow}R$. Except for the region of $r<<1$ which is the vicinity of the  singularity, in most cases $\mathcal{R}$ is rather small and can be safely neglected, thus we will ignore this effect in the rest of this section.
\begin{figure}[htbp]
\centering
\includegraphics[width=0.5\textwidth]{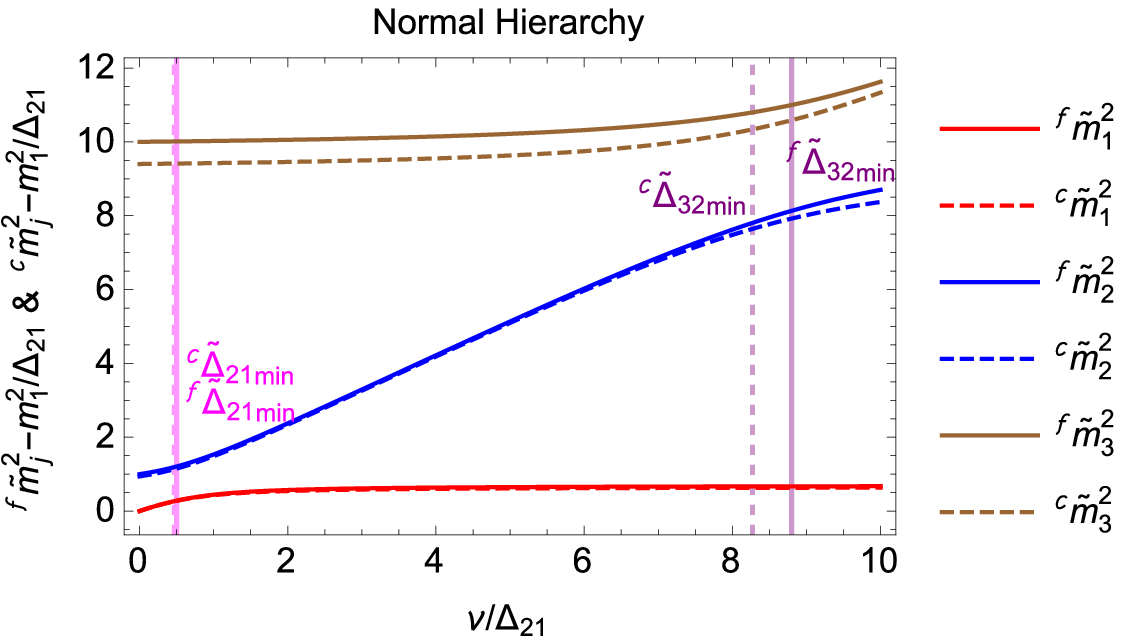}
\includegraphics[width=0.5\textwidth]{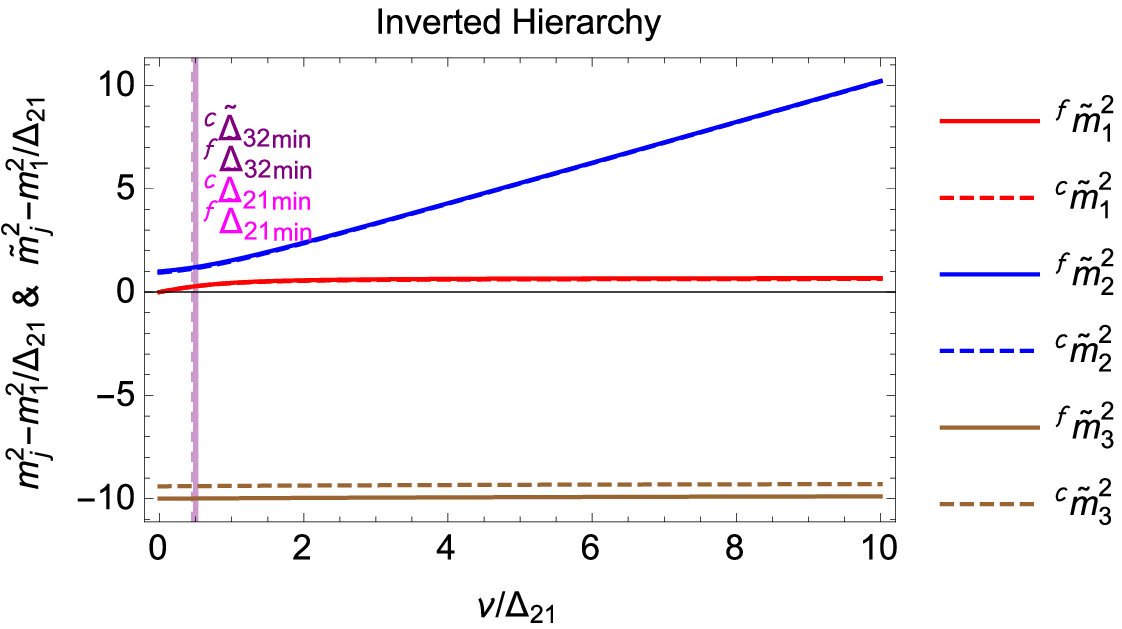}
\caption{Mass eigenvalues in flat and curved spacetime, respectively. Above: in NH. Below: in IH. Red, blue and brown solid (dashed) curves denote the mass eigenvalues $^f \tilde{m}_j^2$ ($^c \tilde{m}_j^2$) for $j=1,2,3$ in flat (curved) spacetime, respectively. $f$ and $c$ denote flat spacetime and curved spacetime, respectively. Solid (dashed) pink vertical lines denote the positions of the minimum of ${^f \tilde{{\Delta}}}_{21}$ ($^c {\tilde{\mathrm{\Delta}}}_{21}$).  Solid (dashed) purple vertical lines are defined similarly except they are for $^f {\tilde{{\Delta}}}_{32}$ ($^c {\tilde{{\Delta}}}_{32}$) instead. Mixing angles are chosen to be $\theta_{12}$=33.48$^\circ$ and $\theta_{13}$=8.52$^\circ$\cite{gonzalezgarcia2012}. Metric is chosen so that $e^{2\Phi}=0.94$ and $e^{2\Lambda}=1$. For illustration purpose, we have set $\Delta_{21}=1$ and $\Delta_{31}=\pm10$ for NH (IH).}\label{fig3}
\end{figure}

Then it is of interest to investigate the influence imposed on the effective masses of neutrinos by the curved spacetime. The dependence of these masses on $v$ is shown in Fig.(\ref{fig3}). It can be seen that the
redshift which is denoted by the deviation of the vertical dashed lines from the solid lines in the figure. At the minimum of $ \tilde{\Delta}_{32}$ the redshift is only obvious
in NH. Whereas, for IH at the minima of  $\tilde{\Delta}_{32}$ and  $\tilde{\Delta}_{21}$
the environmental effect $v$ is almost the same.

Finally we would like to see the overall effect of the
survival probability which may be measurable. Assuming that electron neutrinos are produced in
the core of this celestial object or traversing it, and then departing from the region of the object to
propagate towards the earth. If theses neutrinos could be measured
right after they traverse through the object the overall survival probability is
\begin{gather}
{\mathcal{Q}}_{ee}\mathrm{{=}}{\sin}^{4}{\tilde{\mathit{\theta}}}_{\mathrm{13}}\mathrm{{+}}{\cos}^{4}{\tilde{\mathit{\theta}}}_{\mathrm{13}}{\mathrm{(}}{\cos}^{4}{\tilde{\mathit{\theta}}}_{12}\mathrm{{+}}{\sin}^{4}{\tilde{\mathit{\theta}}}_{12}{\mathrm{)}}.\label{survivalPonlyS}
\end{gather}
\begin{figure}[htbp]
\centering
\includegraphics[width=0.5\textwidth]{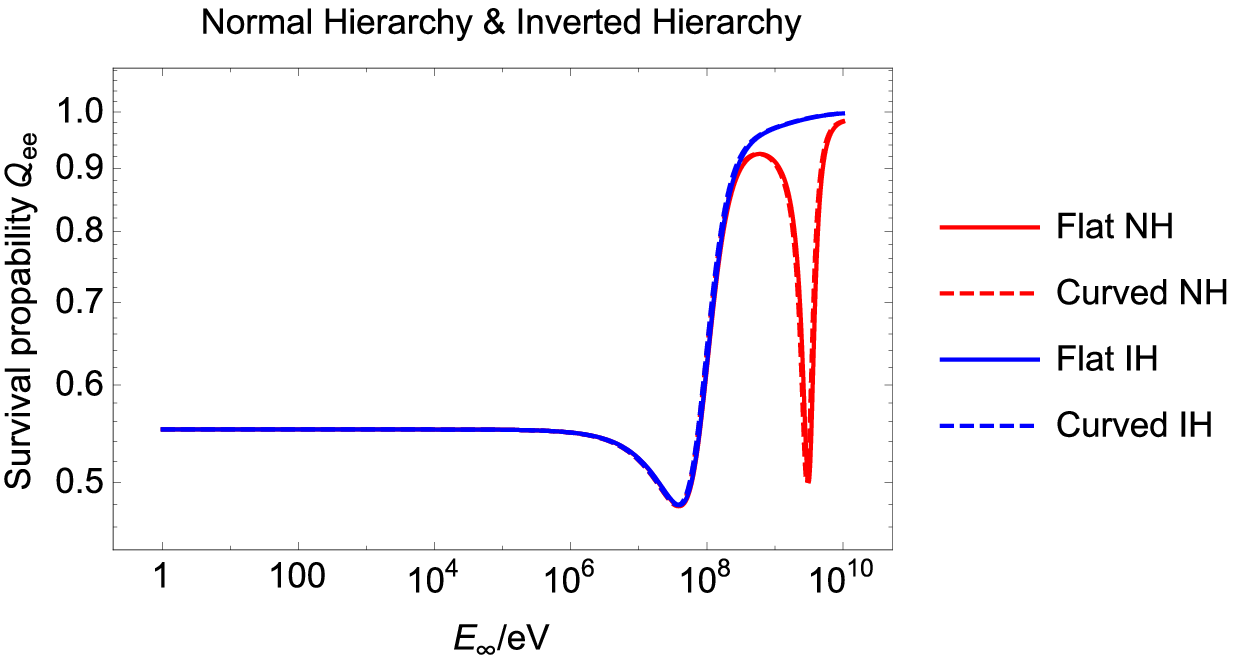}
\includegraphics[width=0.5\textwidth]{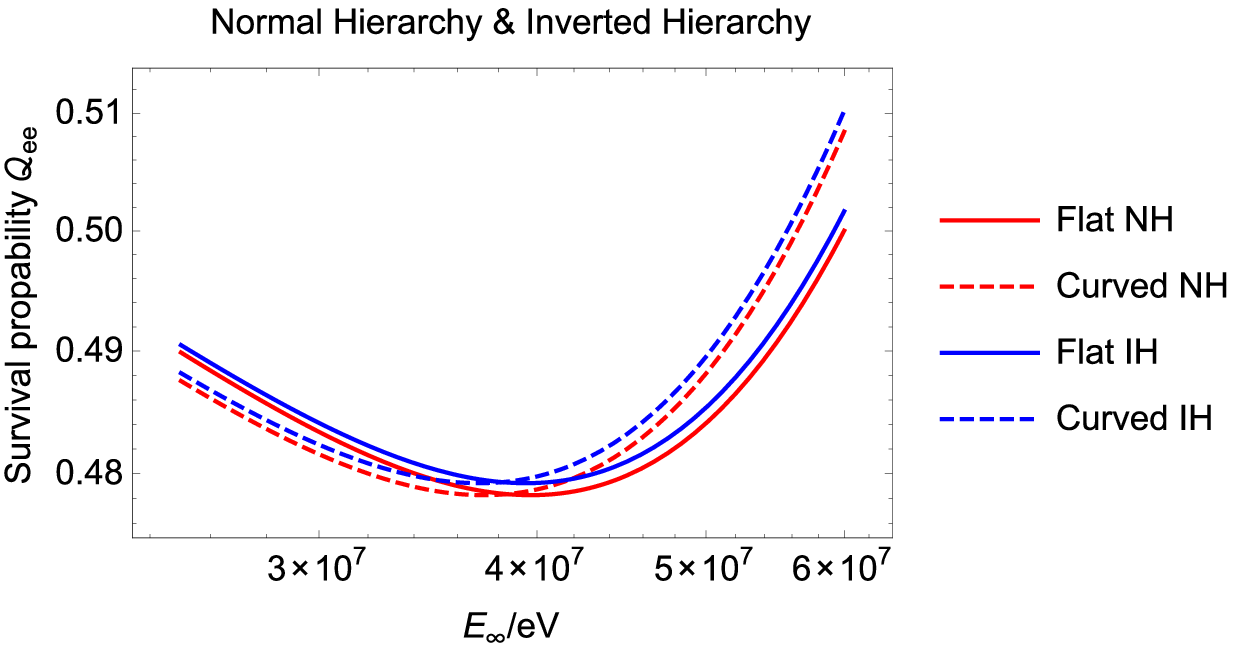}
\caption{Survival probability after traveling through the celestial object. Above: plot range from 1 to $10^{10}$ eV. Below: details around the lower energy (10 MeV) oscillation region. Red and blue solid (dashed) curves denote the survival probability ${\mathcal{Q}}_{ee}$ in flat (curved) spacetime, respectively.  Mixing parameters are chosen as $\theta_{12}$=33.48$^\circ$, $\theta_{13}$=8.52$^\circ$, $\theta_{23}$=42.2$^\circ$, ${\mathrm{\Delta}}_{\mathrm{21}}\mathrm{{=}}{7}{\mathrm{.}}{5}\mathrm{\times}{\mathrm{10}}^{\mathrm{{-}}{5}}$ eV;  ${\mathrm{\Delta}}_{31}\mathrm{{=}}{2}{\mathrm{.}}{5}\mathrm{\times}{\mathrm{10}}^{\mathrm{{-}}{3}}$ eV for NH and ${\mathrm{\Delta}}_{31}\mathrm{{=}}-{2}{\mathrm{.}}{5}\mathrm{\times}{\mathrm{10}}^{\mathrm{{-}}{3}}$ eV for IH\cite{gonzalezgarcia2012}. The metric is chosen as $e^{2\Phi}=0.94$ and $e^{2\Lambda}=1$.}\label{fig4}
\end{figure}
Numerical result of Eq.(\ref{survivalPonlyS}) is plotted in
Fig.(\ref{fig4}). There the gravitational background does not exert any noticeable effect on the overall oscillation behaviors,
but from the figure one can notice a tiny redshift of the resonance energies.

For higher energies, there can be a second resonance dip, while the first one occurs at
about a few of tens MeV which is the characteristic energy of solar and supernova neutrinos. The second dip appears  at about
GeV scale and can only take place in NH. Except for this resonance, the mass
hierarchy plays a limited role in the mixing and the probability
result, thus requires experiments of very high precision to detect.

After the neutrino beam leaves the massive celestial object, it is supposed to freely propagate
towards the earth and will be detected by the earth detectors, then the later stage oscillations are
completely the regular oscillations in vacuum.  Now let us consider the resultant flavor survival probability
determined by the  oscillations in the matter and the sequent vacuum. The survival probability
is then given by
\begin{gather}
{\mathcal{Q}}_{ee}\mathrm{{=}}{c}_{\mathrm{12}}^{2}{\tilde{c}}_{\mathrm{12}}^{2}{c}_{13}^{2}{\tilde{c}}_{13}^{2}\mathrm{{+}}{c}_{13}^{2}{\tilde{c}}_{13}^{2}{s}_{\mathrm{12}}^{2}{\tilde{s}}_{\mathrm{12}}^{2}\mathrm{{+}}{s}_{13}^{2}{\tilde{s}}_{13}^{2},
\label{survivalP}
\end{gather}
\begin{figure}[htbp]
\centering
\includegraphics[width=0.5\textwidth]{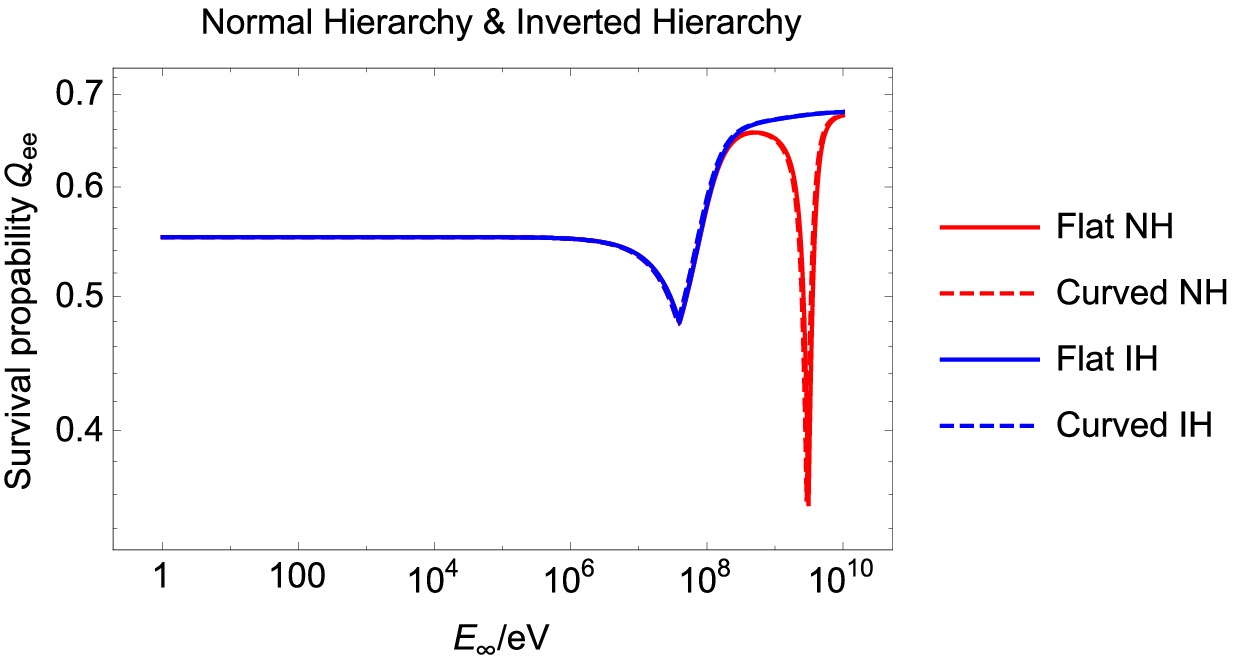}

\includegraphics[width=0.5\textwidth]{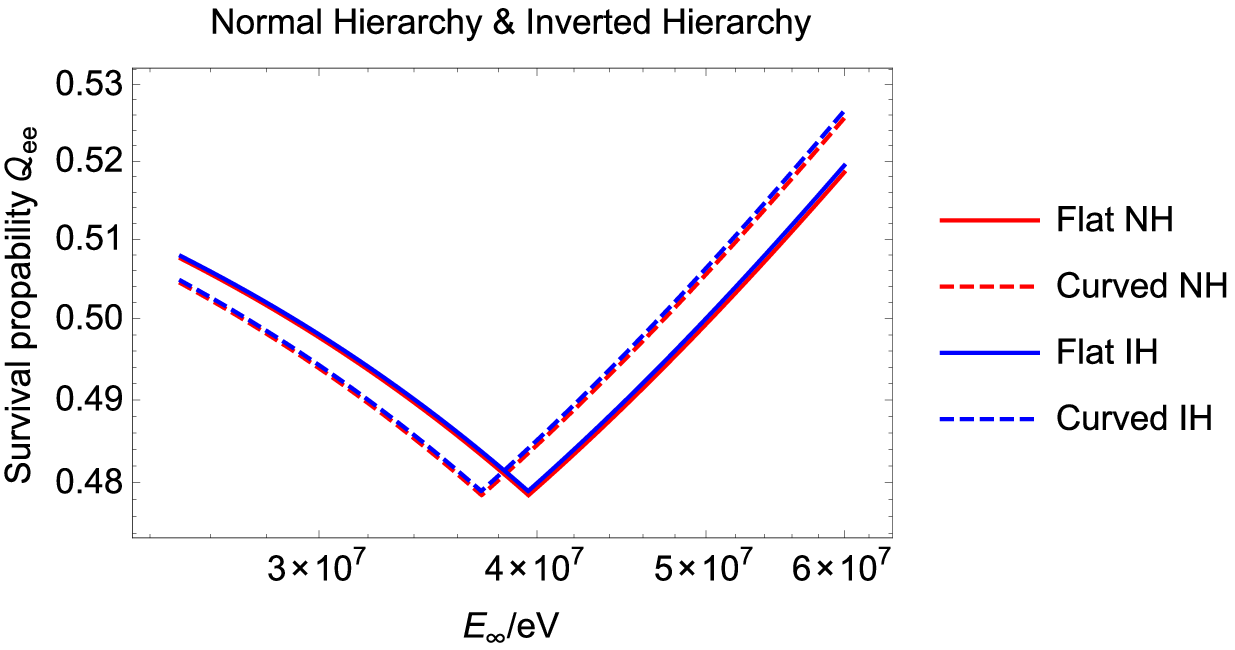}
\caption{Survival probability after traveling through the object and the distance between the object and the earth. Above: plot range from 1 to $10^{10}$ eV. Below: details around the lower energy (10 MeV) oscillation region. Red and blue solid (dashed) curves denote the survival probability ${\mathcal{Q}}_{ee}$ in flat (curved) spacetime, respectively.  Mixing parameters are chosen as $\theta_{12}$=33.48$^\circ$, $\theta_{13}$=8.52$^\circ$, $\theta_{23}$=42.2$^\circ$, ${\mathrm{\Delta}}_{\mathrm{21}}\mathrm{{=}}{7}{\mathrm{.}}{5}\mathrm{\times}{\mathrm{10}}^{\mathrm{{-}}{5}}$ eV, ${\mathrm{\Delta}}_{31}\mathrm{{=}}{2}{\mathrm{.}}{5}\mathrm{\times}{\mathrm{10}}^{\mathrm{{-}}{3}}$   for NH and ${\mathrm{\Delta}}_{31}\mathrm{{=}}-{2}{\mathrm{.}}{5}\mathrm{\times}{\mathrm{10}}^{\mathrm{{-}}{3}}$ eV for IH\cite{gonzalezgarcia2012}. Metric is chosen as $e^{2\Phi}=0.94$ and $e^{2\Lambda}=1$,}\label{fig5}
\end{figure}
the result of which is depicted in Fig.(\ref{fig5}). The spacetime still
mainly influences the resonance energy with a gravitational redshift.

The resonant result indeed imposes an initial condition to the later vacuum oscillation. We notice that under this initial condition, for the lower energy neutrinos (about 10 MeV or below) the vacuum propagation barely changes their behaviors, therefore the propagation of the beam from the production core to the earth is only affected by existence of the
celestial matter. But by contrary, for the higher energy neutrinos (at GeV scale), the probability is seriously modified. The survival probability at this dip goes down from 0.5 to 0.3 and the upper limit of the survival probability could be suppressed from 1 to 0.7.

Keeping the density similar to that of the sun, the dependence of
mass and radius on $e^{2\Phi}$ is shown in Table (\ref{tab1}).
Actually, in our case, if taking $e^{2\Phi}=0.94$, then the energy can
be shifted by 6\% at most. With small mass of the celestial object the redshift would be
very small. As the mass increases, the overall redshift  increases quickly till at
$M=10^8{M}_{\mathrm{\odot}}$, $R=10^{8/3}{R}_{\mathrm{\odot}}$ there
is ${R}\approx{R}_{s}$, where ${R}_{s}$ is the
Schwarzschild radius and ${R}_{s}=\frac{2GM}{{c}^{2}}$.
In this extreme condition the interior Schwarzschild solution we are
using fails and the object collapses, yielding an unreasonable result.

\begin{table}[htbp]
\centering
\begin{tabular}{lll}
\toprule
$M$ & $R$ & $e^{2\Phi}$\\
\midrule
${M}_{\mathrm{\odot}}$ & ${R}_{\mathrm{\odot}}$ & 0.99999\\
$10{M}_{\mathrm{\odot}}$ & ${\mathrm{10}}^{1\mathrm{/}3}{R}_{\mathrm{\odot}}$ & 0.99997\\
$10^3{M}_{\mathrm{\odot}}$ & $10{R}_{\mathrm{\odot}}$ & 0.99936\\
$10^6{M}_{\mathrm{\odot}}$ & $10^2{R}_{\mathrm{\odot}}$ & 0.93678\\
$10^7{M}_{\mathrm{\odot}}$ & $10^{7/3}{R}_{\mathrm{\odot}}$ & 0.71304\\
$10^8{M}_{\mathrm{\odot}}$ & $10^{8/3}{R}_{\mathrm{\odot}}$ & 0.00329\\
\bottomrule
\end{tabular}
\caption{Redshift factor $e^{2\Phi}$ vs. the mass and radius of the object}\label{tab1}
\end{table}
This redshift  will be much more manifest in supernovae where the
density of matter can be as much as $\mathrm{\sim}{\mathrm{10}}^{\mathrm{13}}$
times larger than the solar density. The reason why we stick to
choosing the solar density is that the MSW effect in the sun has been well experimentally and
theoretically confirmed, so that we have something solid to help in making sense. For the influence of density on the energy of the first oscillation dip, see Table (\ref{tab2}). It should be clarified that
this work is only to set an upper bound on the gravitational
modifications to the MSW effect. The real gravitational effect may be somehow different.
\begin{table}[htbp]
\centering
\begin{tabular}{llllll}
\toprule
$\rho/{\mathit{\rho}}_{\mathrm{\odot}}$& $10$ & $20$ & $30$ & $40$& $50$\\
\midrule
${\left.{{E}_{\mathrm{\infty}}}\right|}_{res}/$eV & ${3}{\mathrm{.}}{9}\mathrm{\times}{\mathrm{10}}^{6}$ & ${2}{\mathrm{.}}{0}\mathrm{\times}{\mathrm{10}}^{6}$ & ${1}{\mathrm{.}}{2}\mathrm{\times}{\mathrm{10}}^{6}$ & ${9}{\mathrm{.}}{8}\mathrm{\times}{\mathrm{10}}^{5}$& ${7}{\mathrm{.}}{2}\mathrm{\times}{\mathrm{10}}^{5}$\\
\bottomrule
\end{tabular}
\caption{Energy of the first resonance dip vs. density of the celestial object. The mass of the object is selected to be 10${M}_{\mathrm{\odot}}.$}\label{tab2}
\end{table}
\subsection{Varying density}
Now let us turn to the more realistic situation in which the density profile of the celestial object is arbitrarily varying. In this scheme the neutrino is produced inside the object, or just travels through the object, then leaves the object and propagates towards the Earth and finally is received by detectors on the ground. The varying density of the celestial object renders the mixing parameters dependent on the position and time, i.e., ${\tilde{\mathit{\theta}}}_{ij}\mathrm{\rightarrow}{\tilde{\mathit{\theta}}}_{ij}{\mathrm{(}}\mathit{\rho}{\mathrm{(}}{r}{\mathrm{,}}{t}{\mathrm{))}}$. This composes a new degree of freedom and induces a new effect: the change of admixtures, meaning that the mass eigenstates $\left.{\left|{{\mathit{\nu}}_{j}}\right.}\right\rangle$ are not eigenstates of propagation anymore and dependent on position and time. Furthermore, for a neutrino of certain energy, inside the celestial object there will exist a (possibly narrow) resonance layer where $
\mathit{\rho}{\mathrm{(}}{r}{\mathrm{,}}{t}{\mathrm{)}}\mathrm{{=}}{\mathit{\rho}}_{R}$, with $
{\mathit{\rho}}_{R}
$
being the resonance density.

Excluding dynamics of the medium, the density is only dependent on the position. Under these circumstances the survival probability becomes
\begin{gather}
{\mathcal{Q}}_{ee}\mathrm{{=}}{c}_{\mathrm{13}}^{2}{\tilde{c}}_{\mathrm{13}}^{2}{\mathrm{(}}{r}_{0}{\mathrm{)(}}\frac{1}{2}\mathrm{{+}}{\mathrm{(}}\frac{1}{2}\mathrm{{-}}{P}_{C}{\mathrm{)}}\cos{2}{\mathit{\theta}}_{\mathrm{12}}\cos{2}{\tilde{\mathit{\theta}}}_{\mathrm{12}}{\mathrm{(}}{r}_{0}{\mathrm{))}}\mathrm{{+}}{s}_{\mathrm{13}}^{2}{\tilde{s}}_{\mathrm{13}}^{2}{\mathrm{(}}{r}_{0}{\mathrm{)}},
\end{gather}
where ${P}_{C}$ stands for the ${\mathit{\nu}}_{1}\mathrm{/}{\mathit{\nu}}_{2}$ level crossing probability and $r_0$ is the radial coordinate of the initial position of neutrino.


To calculate this survival probability, one still needs to find the mixing parameters at the initial point and the metric of background spacetime. Yet if taking the density to be non-constant, Eq.(\ref{ce1}) and (\ref{interiorS}) fail and one will have to solve differential equations Eq.(\ref{evolutionE}) (evolution equation) or Eq.(\ref{mixingE}) (mixing equation) and the Oppenheimer-Volkoff equation in order to obtain the components of each flavor or mixing parameters and the metric for the spacetime. In general it would be difficult solving them analytically even for linear or exponential density profiles. For a specific problem, it is usually only solvable via numerical methods and the results are model-dependent.

What is more, because of the self-energy correction by gravitational filed discussed earlier in Sec.(\ref{s3}), the effective potential of neutrino acquires a dependence on the spacetime. With this new degree of freedom there will be new effects. The oscillation condition of MSW effect will be switched as
\begin{gather}
\mathit{\rho}{\mathrm{(}}{r}{\mathrm{)}}\mathrm{{=}}{\mathit{\rho}}_{R}\mathrm{\rightarrow}\mathit{\rho}{\mathrm{(}}{r}{\mathrm{)}}\mathrm{{=}}{\mathit{\rho}}_{R}{\mathrm{(}}{r}{\mathrm{)}},
\end{gather}
which means the resonance density ${\mathit{\rho}}_{R}$ is not constant anymore and there might be zero, one, or even multiple resonance layers in the objects, depending on specific structure of density profiles $\mathit{\rho}{\mathrm{(}}{r}{\mathrm{)}}$ and the spacetime. Surely these effects may be fairly small if not at the vicinities of spacetime singularities.

\section{Discussion}\label{s7}
In this work we review the geometric treatment for neutrino
oscillations in curved spacetime and then extend it to the three-generation cases.
A discussion of the gravitational effect on neutrino
self-energy is given and the corresponding correction to the neutrino effective
potential is derived. Applying this method and the results, survival probabilities of
oscillations in vacuum and the MSW effect in a general, static and
spherically symmetric spacetime for the three-generation neutrino scenario
are then calculated. For the matter effect in curved spacetime, mixing
parameters of neutrinos are evaluated for celestial objects with constant densities, and both
normal and inverted mass hierarchy cases are respectively discussed. In the end of
this work, numerical results  for the background of a static,
isotropic interior Schwarzschild spacetime are shown in figures.

The  modification due to the gravitational background to the neutrino self-energy Eq.(\ref{selfenergy})
implies an overall energy shift that is induced by the tetrad and an extra
gravitational phase, however the dependence of the phase on energy is
not separable from the common matter effect. The reason is that this phase has the same dependence on $G_F$, $N_e$ and $E_\infty$ as the original MSW effect.
Thus this extra phase caused by the gravitational background and the original MSW phase would mix together and become
indistinguishable at the energy spectrum, resulting in an overall redshift, even though formally this geometric-induced extra phase can be separated
from the original propagation phase. Furthermore, this new
effect is
$\mathrm{{-}}\mathrm{\frac{\mathcal{R}{\hbar}^{2}}{16{m}_{e}^{2}{c}^{2}}}$,
which is proportional to the Ricci scalar curvature $\mathcal{R}$ of
the spacetime and therefore, it will vanish in the Schwarzschild spacetime,
since in the Schwarzschild metric $\mathcal{R}$=0 and it describes a vacuum spacetime. Also, for this effect to be manifest we need that
${\mathcal{R}}\mathrm{\sim}\frac{{m}_{e}^{2}{c}^{2}}{{\mathrm{\hbar}}^{2}}\mathrm{\sim}{\mathrm{10}}^{\mathrm{24}}$
m$^{-2}$, which is dramatically large and only possible at the
vicinities of spacetime singularities.

The three-generation vacuum oscillations in curved spacetime have
exactly the same form as their two-generation situation and all the
conclusions can be generalized directly. On the other hand, the formulation for the MSW
effect with three generations in curved spacetime becomes rather
lengthy and require a separate discussion for normal hierarchy and
inverted hierarchy. The major influence of gravitation is the energy
redshift caused by a factor $
{\mathrm{(}}{1}\mathrm{{-}}\mathrm{\frac{\mathcal{R}{\hbar}^{2}}{16{m}_{e}^{2}{c}^{2}}}{\mathrm{)}}^{-1}{e}^{\mathrm{2\Phi}}
$. The numerical results are then given in two scenarios; matter effect
only and the overall survival probability including the matter effect and the later stage vacuum
oscillations.

The extra resonance dip of higher energy only exists in the normal mass hierarchy. Therefore detecting it can help solve the mass hierarchy problem. Besides this significant difference, the mass hierarchy also plays a role in determining the survival probability curve of neutrinos and the dip at lower energy, even though detecting
such effect is rather difficult, if not impossible in the future.

For astronomical objects with varying densities, a method of calculation is offered. In curved spacetime there might exist zero, one or even multiple resonance layers, though these effects require high-precision experiments in order to detect, but actually they may become significant only at the vicinities of spacetime singularities. At the present stage it would be rather difficult to acquire any detailed information about structures and density profiles of those celestial objects. But with these neutrino experiments we can expect to learn the structures of these celestial objects.

In 2016 the gravitational wave  emitted from a binary
black hole merger\cite{abbott2016} was eventually discovered by the LIGO collaboration. One can immediately conjecture that besides the
gravitational wave emission, the electromagnetic radiation and neutrino emission would also take place just like the supernova 1987A.
Events at such scale, the
gravitational field must play a role to induce local quantum effects and produce a great amount of neutrinos. The photons might be
scattered away or absorbed by the celestial objects on the long journey to our earth which is much longer than the distance from Supernova 1987A to the earth, but the neutrinos should come along with the gravitational wave. So far, because of
lacking detailed information, we cannot determine the neutrino energy spectra yet.
Even though the IceCube and ANTARES collaborations
reported that they did not
detect any neutrino excess accompanying the observation of gravitational wave \cite{IceCube}, it is really natural
to believe that there is no reason to prohibit neutrino burst along with such events, so we should have received some extra neutrinos.
Therefore, one may guess that due to
unknown reasons, the neutrinos produced in the black hole merger disappear or somehow evade our detection. The mechanism is
worth careful studies and the effects of the gravitational field which deforms the flat spacetime to curved are also needed to
investigate. Our recent work is only a step towards the aim and we indeed find the curve spacetime can influence the
neutrino propagation. We would like to emphasize that we have no intention to draw a comparison between the energy spectra or production mechanisms of the supernova neutrinos and that of the neutrinos from  binary black hole merger. The two sources have completely the different strengths of gravitational fields, thus the possibly observable effects would be different. We are discussing the scenarios and wish to draw researchers' attention to the gravitational effect on neutrino oscillation. On other aspects,
because  many approximations are adopted in the study,
the obtained results are not complete, more work is badly needed for getting a better understanding of mysteries of the cosmic neutrinos.
\section{Acknowledgement}
This work is supported by the National Natural Science Foundation of China under contract
No.11375128 and 11135009.
\newpage
\section*{Appendix}\label{appen}
In this appendix we review the geometric treatment of neutrino phase differences\cite{cardall97}.

Any wave function of neutrino flavor eigenstates can be expanded in the space of mass eigenstates
\begin{gather}
\left.{\left|{{\mathrm{\Psi}}_{\mathit{\alpha}}\mathrm{(}x\mathrm{,}t\mathrm{)}}\right.}\right\rangle\mathrm{{=}}\mathop{\sum}\limits_{j}{{U}_{\mathit{\alpha}{j}}}{T}^{\mathrm{\dag}}\left.{\left|{{\mathit{\nu}}_{j}}\right.}\right\rangle, \label{equ1}
\end{gather}
where $\mathit{\alpha}\mathrm{{=}}\mathit{e}{\mathrm{,}}\mathit{\mu}{\mathrm{, }}\mathit{\tau}$ represents the three neutrino flavor eigenstates and ${j}\mathrm{{=}}{1}{\mathrm{,}}{2}{\mathrm{,}}{3}$ reprensents the three mass eigenstates. ${U}_{ij}$ is the element of the PMNS matrix and $
{T}^{\mathrm{\dag}}
$ is the spacetime evolution operator.
Neutrino mass states are free solutions to the corresponding Dirac equations, thus they can be expressed as plane waves with the Minkowski metric being chosen as $diag(-1,1,1,1)$
\begin{gather}
{T}^{\mathrm{\dag}}\left.{\left|{{\mathit{\nu}}_{j}}\right.}\right\rangle\mathrm{{=}}{e}^{{iP}^{\mathit{\mu}}{x}_{\mathit{\mu}}}\left.{\left|{{\mathit{\nu}}_{j}}\right.}\right\rangle, \label{equ2}
\end{gather}
where ${P}^{\mathit{\mu}}\mathrm{{=}}{\mathrm{(}}{E}{\mathrm{,}}{P}_{j}^{1}{\mathrm{,}}{0}{\mathrm{,}}{0}{\mathrm{)}}$ is the four-momentum of the neutrino whose trajectory is assumed to be null geodesics. The mass-shell condition is ${P}_{j}^{1}\mathrm{\approx}{E}\mathrm{{-}}\frac{{M}_{j}^2}{2E}$, where $M_j$ is the mass operator that yields the mass of neutrino when acts on a neutrino state.


In curved spacetime where the neutrino travels along a null geodesic line instead of a straight line, Eq.(\ref{equ2}) can be re-written as
\begin{gather}
{T}^{\mathrm{\dag}}\left.{\left|{{\mathit{\nu}}_{j}}\right.}\right\rangle\mathrm{{=}}{e}^{{i}\int{{P}^{\mathit{\mu}}{p}_{\mathit{\mu}}{d}\mathit{\tau}}}\left.{\left|{{\mathit{\nu}}_{j}}\right.}\right\rangle, \label{equ3}
\end{gather}
where the tangent vector to the null geodesic line on which the neutrino travels is defined to be ${p}^{\mathit{\mu}}\mathrm{{=}}\frac{{dx}^{\mathit{\mu}}}{{d}\mathit{\tau}}$ with $\mathit{\tau}$ an affine parameter.

After modifications of wave functions, in curved spacetime the translation of Dirac equation is also needed. On a four dimensional, torsion-free spacetime depicted by a pseudo-Riemann manifold $\mathcal{M}$, the Dirac equation in electron environment with metric sign ${\mathrm{(}}\mathrm{{-}}\mathrm{{+}}\mathrm{{+}}\mathrm{{+}}{\mathrm{)}}$
reads
\begin{gather}
{\mathit{\gamma}}^{\mathit{\mu}}{\mathrm{(}}{i}{\mathcal{D}}_{\mathit{\mu}}\mathrm{{-}}{A}_{\mathit{\mu}}{\mathrm{)}}\mathrm{\Psi}\mathrm{{+}}{M}\mathrm{\Psi}\mathrm{{=}}{0},\label{dirac}
\end{gather}
where ${A}_{\mu}$ denotes the effective potential caused by the electron environment, the covariant derivative ${\mathcal{D}}_{\mathit{\mu}}$ and Dirac matrices ${\mathit{\gamma}}^{\mathit{\mu}}$ on the manifold are defined as
\begin{gather}
{\mathcal{D}}_{\mathit{\mu}}\mathrm{{=}}{\mathrm{\partial}}_{\mathit{\mu}}\mathrm{{+}}{\mathrm{\Gamma}}_{\mathit{\mu}},\\
{\mathit{\gamma}}^{\mathit{\mu}}\mathrm{{=}}{e}_{a}^{\mathit{\mu}}{\mathit{\gamma}}^{a}\label{gamma}.
\end{gather}
The spinor connection ${\mathrm{\Gamma}}_{\mathit{\mu}}$, spin connection ${\mathit{\omega}}_{{ab}\mathit{\mu}}$ and tetrad ${e}_{a}^{\mathit{\mu}}$ are defined to be
\begin{gather}
{\mathrm{\Gamma}}_{\mathit{\mu}}\mathrm{{=}}\frac{1}{8}{\mathit{\omega}}_{{ab}\mathit{\mu}}{\mathrm{[}}{\mathit{\gamma}}^{a}{\mathrm{,}}{\mathit{\gamma}}^{b}{\mathrm{]}},\\
{\mathit{\omega}}_{{ab}\mathit{\mu}}\mathrm{{=}}{e}_{a}^{\mathit{\nu}}{\mathrm{\partial}}_{\mathit{\mu}}{e}_{{b}\mathit{\nu}}\mathrm{{-}}{e}_{a}^{\mathit{\nu}}{\mathrm{\Gamma}}_{\mathit{\mu}\mathit{\nu}}^{\mathit{\sigma}}{e}_{{b}\mathit{\sigma}},\\
{g}_{\mathit{\mu}\mathit{\nu}}{e}_{a}^{\mathit{\mu}}{e}_{b}^{\mathit{\nu}}\mathrm{{=}}{\mathit{\eta}}_{ab}.
\end{gather}


The mass shell condition (Hamilton-Jacobi equation) for neutrinos can be obtained from Eq.(\ref{dirac}) as
\begin{gather}
{\mathrm{(}}{P}^{\mathit{\mu}}\mathrm{{+}}{A}^{\mathit{\mu}}{\mathrm{)(}}{P}_{\mathit{\mu}}\mathrm{{+}}{A}_{\mathit{\mu}}{\mathrm{)}}\mathrm{{=}}\mathrm{{-}}{M}^{2}.\label{equpp}
\end{gather}

Since the the trajectories are null geodesics where the proper time vanishes, we cannot simply use the proper time as the affine parameter to build the relation between the four-momentum and the tangent vector. But we can still demand that ${P}^{0}\mathrm{{=}}{p}^{0}$ and the rest three spatial components parallel to each other. Then from Eq.(\ref{equpp}) we obtain
\begin{gather}
{P}^{\mathit{\mu}}{p}_{\mathit{\mu}}\mathrm{{=}}\mathrm{{-}}{\mathrm{(}}\frac{{M}^{2}}{2}\mathrm{{+}}{p}^{\mathit{\mu}}{A}_{\mathit{\mu}}{\mathrm{)}}.\label{Pp}
\end{gather}
Combining Eq.(\ref{equ1}), Eq.(\ref{equ3}) and Eq.(\ref{Pp}) we finally obtain the propagation of a flavor state in spacetime
\begin{gather}
\left.{\left|{{\mathrm{\Psi}}_{\mathit{\alpha}}{\mathrm{(}}\mathit{\tau}{\mathrm{)}}}\right.}\right\rangle\mathrm{{=}}\mathop{\sum}\limits_{j}{{U}_{\mathit{\alpha}{j}}{e}^{i\int{{\mathrm{(}}\frac{{M}^{2}}{2}\mathrm{{+}}{p}^{\mathit{\mu}}{A}_{\mathit{\mu}}{\mathrm{)}}{d}\mathit{\tau}}}}\left.{\left|{{\mathit{\nu}}_{j}}\right.}\right\rangle. \label{propagation}
\end{gather}
To calculate Eq.(\ref{propagation}), we replace the integral element with a proper length $dl$ defined as
\begin{gather}
{dl}^{2}\mathrm{\equiv}{g}_{ij}{dx}^{i}{dx}^{j}.\label{pl}
\end{gather}
Taking square root and dividing the affine parameter into it yields
\begin{gather}
{dl}\mathrm{{=}}{\mathrm{(}}{g}_{ij}{dx}^{i}{dx}^{j}{\mathrm{)}}^{1\mathrm{/}2}\mathrm{{=}}{\mathrm{(}}{g}_{ij}\frac{{dx}^{i}}{{d}\mathit{\tau}}\frac{{dx}^{j}}{{d}\mathit{\tau}}{\mathrm{)}}^{1\mathrm{/}2}{d}\mathit{\tau}. \label{geode}
\end{gather}
The integral element can be obtained as
\begin{gather}
{d}\mathit{\tau}\mathrm{{=}}{dl}{\mathrm{(}}\mathrm{{-}}{g}_{\mathrm{00}}{\mathrm{(}}\frac{{dx}^{0}}{{d}\mathit{\tau}}{\mathrm{)}}^{2}{\mathrm{)}}^{\mathrm{{-}}{1}{\mathrm{/}}{2}}.\label{geode2}
\end{gather}
From Eq.(\ref{geode}) to Eq.(\ref{geode2}) we have applied the geodesic equation $
{g}_{\mathit{\mu}\mathit{\nu}}\frac{{dx}^{\mathit{\mu}}}{{d}\mathit{\tau}}\frac{{dx}^{\mathit{\nu}}}{{d}\mathit{\tau}}\mathrm{{=}}{0}\mathrm{\Rightarrow}{g}_{ij}\frac{{dx}^{i}}{{d}\mathit{\tau}}\frac{{dx}^{j}}{{d}\mathit{\tau}}\mathrm{{=}}\mathrm{{-}}{g}_{\mathrm{00}}{\mathrm{(}}\frac{{dx}^{0}}{{d}\mathit{\tau}}{\mathrm{)}}^{2}
$.


\end{document}